\documentstyle[11pt,amssymb]{article}

\makeatletter
\@addtoreset{equation}{section}
\makeatother

\def\crampest{\medmuskip = 1mu plus 1mu minus 1mu}
\def\uncramp{\medmuskip = 4mu plus 2mu minus 4mu}
\def\ben{\begin{equation}}
\def\een{\end{equation}}

\let\a=\alpha    
    
  \let\n=\nu   
 \let\t=\tau

\let\C=\Chi

\def\nn{\nonumber} \def\bd{\begin{document}} \def\ed{\end{document}}
\def\ds{\documentstyle} \let\fr=\frac \let\bl=\bigl \let\br=\bigr
\let\Br=\Bigr \let\Bl=\Bigl
\let\bm=\bibitem
\let\na=\nabla
\let\pa=\partial \let\ov=\overline
\newcommand{\be}{\begin{equation}}
\newcommand{\ee}{\end{equation}}
\def\ba{\begin{array}}
\def\ea{\end{array}}
\def\ft#1#2{{\textstyle{{\scriptstyle #1}\over {\scriptstyle #2}}}}
\def\fft#1#2{{#1 \over #2}}
\def\del{\partial}
\def\vp{\varphi}
\def\sst#1{{\scriptscriptstyle #1}}
\def\oneone{\rlap 1\mkern4mu{\rm l}}
\def\td{\tilde}
\def\wtd{\widetilde}
\def\ie{\rm i.e.\ }
\def\dalemb#1#2{{\vbox{\hrule height .#2pt
        \hbox{\vrule width.#2pt height#1pt \kern#1pt
                \vrule width.#2pt}
        \hrule height.#2pt}}}
\def\square{\mathord{\dalemb{6.8}{7}\hbox{\hskip1pt}}}
\newcommand{\ho}[1]{$\, ^{#1}$}
\newcommand{\hoch}[1]{$\, ^{#1}$}
\newcommand{\bea}{\begin{eqnarray}}
\newcommand{\eea}{\end{eqnarray}}
\newcommand{\ra}{\rightarrow}
\newcommand{\lra}{\longrightarrow}
\newcommand{\Lra}{\Leftrightarrow}
\newcommand{\ap}{\alpha^\prime}
\newcommand{\bp}{\tilde \beta^\prime}
\newcommand{\tr}{{\rm tr} }
\newcommand{\Tr}{{\rm Tr} }
\def\0{{\sst{(0)}}}
\def\1{{\sst{(1)}}}
\def\2{{\sst{(2)}}}
\def\3{{\sst{(3)}}}
\def\4{{\sst{(4)}}}
\def\5{{\sst{(5)}}}
\def\6{{\sst{(6)}}}
\def\7{{\sst{(7)}}}
\def\8{{\sst{(8)}}}
\def\n{{\sst{(n)}}}
\def\cA{{{\cal A}}}
\def\cF{{{\cal F}}}
\def\tV{\widetilde V}
\def\tW{\widetilde W}
\def\tH{\widetilde H}
\def\tE{\widetilde E}
\def\tF{\widetilde F}
\def\tA{\widetilde A}
\def\im{{{\rm i}}}
\def\tY{{{\wtd Y}}}
\def\ep{{\epsilon}}
\def\vep{{\varepsilon}}
\def\R{\rlap{\rm I}\mkern3mu{\rm R}}
\def\bD{{{\bar D}}}
\def\alp{{{\a'}^3}}
\def\R{\rlap{\rm I}\mkern3mu{\rm R}}
\def\bD{{{\bar D}}}
\def\R{{{\Bbb R}}}
\def\C{{{\Bbb C}}}
\def\H{{{\Bbb H}}}
\def\CP{{{\Bbb C}{\Bbb P}}}
\def\RP{{{\Bbb R}{\Bbb P}}}
\def\Z{{{\Bbb Z}}}
\def\bA{{{\Bbb A}}}
\def\bB{{{\Bbb B}}}
\def\bC{{{\Bbb C}}}
\def\bR{{{\Bbb R}}}
\def\bD{{{\Bbb D}}}
\def\bE{{{\Bbb E}}}
\def\bZ{{{\Bbb Z}}}
\def\Re{{{\frak{Re}}}}
\def\Im{{{\frak{Im}}}}
\def\cosec{{\,\hbox{cosec}\,}}
\def\Gm{{\Gamma_{\!\! -}}}
\def\Gp{{\Gamma_{\!\! +}}}
\def\stan{{standard }}
\def\nonstan{{supernumerary }}

\def\cosech{{\hbox{cosech}}}

\def\etcyc{{\hbox{and cyclic}}}
\def\btheta{{\bar\theta}}

\newcommand{\tamphys}{\it Center for Theoretical Physics,
Texas A\&M University, College Station, TX 77843, USA}

\newcommand{\mitchell}{\it George P. \& Cynthia W.
Mitchell Institute for Fundamental Physics,\\
Texas A\&M University, College Station, TX 77843-4242, USA}
\newcommand{\umich}{\it Michigan Center for Theoretical Physics,
University of Michigan\\ Ann Arbor, MI 48109, USA}
\newcommand{\upenn}{\it Department of Physics and Astronomy,
University of Pennsylvania, Philadelphia,  PA 19104, USA}
\newcommand{\SISSA}{\it  SISSA-ISAS and INFN, Sezione di Trieste\\
Via Beirut 2-4, I-34013, Trieste, Italy}

\newcommand{\newton}{\it Isaac Newton Institute for Mathematical Sciences,\\
20 Clarkson Road,  University of Cambridge,
Cambridge CB3 0EH, UK}

\newcommand{\ihp}{\it Institut Henri Poincar\'e\\
  11 rue Pierre et Marie Curie, F 75231 Paris Cedex 05}

\newcommand{\damtp}{\it DAMTP, Centre for Mathematical Sciences,
 Cambridge University\\  Wilberforce Road, Cambridge CB3 OWA, UK}
\newcommand{\itp}{\it Institute for Theoretical Physics, University of
California\\ Santa Barbara, CA 93106, USA}

\newcommand{\imperial}{\it The Blackett Laboratory, Imperial College London\\
Prince Consort Road, London SW7 2AZ. }

\newcommand{\auth}{
H. L\"u\hoch{\ddagger1},
C.N. Pope\hoch{\ddagger1} and K.S. Stelle\hoch{\star2} }

\begin{document}
\begin{flushright}
\hfill{
MIFP-03-01 \ \
Imperial/TP/03-04/2}\\
\hfill{
\bf hep-th/0311018}
\end{flushright} 

\begin{center}  

{\Large {\bf Higher-Order Corrections to
Non-Compact Calabi-Yau Manifolds in String Theory 
}}   

\vspace{15pt}

\auth

\vspace{7pt}
{\hoch{\ddagger}\mitchell}

\vspace{7pt}
{\hoch{\star}\imperial} 

\vspace{30pt}

\underline{ABSTRACT}
\end{center}  

   At the leading order, the low-energy effective field equations in
string theory admit solutions of the form of products of Minkowski
spacetime and a Ricci-flat Calabi-Yau space.  The equations of motion
receive corrections at higher orders in $\alpha'$, which imply that
the Ricci-flat Calabi-Yau space is modified.  In an appropriate choice
of scheme, the Calabi-Yau space remains K\"ahler, but is no longer
Ricci-flat.  We discuss the nature of these corrections at order
${\alpha'}^3$, and consider the deformations of all the known
cohomogeneity one non-compact K\"ahler metrics in six and eight
dimensions.  We do this by deriving the first-order equations
associated with the modified Killing-spinor conditions, and we thereby
obtain the modified supersymmetric solutions.  We also give a detailed
discussion of the boundary terms for the Euler complex in six and
eight dimensions, and apply the results to all the cohomogeneity one
examples.

{\vfill\leftline{}\vfill
\vskip 10pt \footnoterule
{\footnotesize \hoch{1}
Research supported in part by DOE grant DE-FG03-95ER40917
\vskip -12pt} \vskip 14pt
{\footnotesize \hoch{2}
Research supported in part by the EC under TMR
contract HPRN-CT-2000-00131 and by PPARC \\
$\phantom{xxxx}$ under SPG grant
PPA/G/S/1998/00613.
\vskip -12pt}  \vskip  14pt
}

\pagebreak
\setcounter{page}{1}

\tableofcontents
\addtocontents{toc}{\protect\setcounter{tocdepth}{2}}
\newpage

\section{Introduction}

   Calabi-Yau manifolds have played a central r\^ole in string theory,
by providing the compactifying spaces that permit four-dimensional
effective actions to be extracted from ten-dimensional strings, via a
Kaluza-Klein mechanism \cite{cahostwi}.  In this context, the
requisite Calabi-Yau manifolds are six-dimensional, and they must be
compact so that the Kaluza-Klein spectrum will be discrete, with a
mass gap.  The special holonomy, $SU(3)$, of the Calabi-Yau spaces is
a crucial aspect of their structure, since it implies that there will
be $N=1$ supersymmetry in the four-dimensional spacetime.

    More recently, within the framework of the AdS/CFT correspondence,
Calabi-Yau manifolds and other spaces of special holonomy that are
instead {\it non-compact} have found a natural r\^ole.  They can
provide gravity duals for superconformal field theories with less than
the maximal supersymmetry on the boundaries of anti-de Sitter
spacetimes that arise in the decoupling limits of D-branes or
M-branes.

   At leading order, the effective equations of motion in string
theory imply that a configuration of the form (Minkowski)$_d\times
K_{10-d}$ will give a solution if the ``internal'' manifold $K_{10-d}$
is Ricci-flat.  The further requirement of unbroken supersymmetry
implies that it should have special holonomy.  Beyond the leading
order, there are correction terms in the effective action that modify
the equations of motion that the background must satisfy.  In
particular, there are corrections, starting at order ${\a'}^3$, which
imply in general that the internal manifold will no longer be
Ricci-flat.  This is the case even in situations with supersymmetry,
such as when $K$ is a K\"ahler manifold.

   In this paper we study the effects of the ${\a'}^3$ corrections in
detail for several examples of six-dimensional and eight-dimensional
K\"ahler manifolds.  The metrics that we consider are all of
cohomogeneity one, which means that the Einstein equation, together
with the higher-order corrections, gives rise to a system of coupled
ordinary differential equations for metric functions.  At leading
order the metrics are Ricci flat.  Our examples in six dimensions
include the resolved and deformed conifolds, and the $\R^2$ bundle
over $\CP^2$ or $S^2\times S^2$.  In eight dimensions we consider
$\R^2$ bundles over $S^2\times S^2\times S^2$, $S^2\times \CP^2$ or
$\CP^3$; $\R^4$ bundles over $S^2\times S^2$ or $\CP^2$; and the
Stenzel metric on the $\R^4$ bundle over $S^4$.  (The ${\a'}^3$
corrections for the six dimensional resolved and deformed conifolds,
and the $\R^2$ bundle over $S^2\times S^2$, were previously studied in
\cite{tseytlin}.)  In each case we derive first-order systems of
equations that describe the corrections to Ricci-flatness implied by
the ${\a'}^3$ terms in the string effective action.  We obtain a
general implicit solution of the corrected first-order equations, and
then we solve them explicitly in a perturbative approach.  We show how
they lead to non-singular modifications of the original Ricci-flat
metrics.  The perturbative analysis is valid provided that the
string scale $\sqrt{\a'}$ is small compared with the scale-size $L$
of the Calabi-Yau metric.  This scale size is characterised by the 
size of the bolt at short distance.

   Our analysis can easily be extended to include corrections at
order higher than ${\a'}^3$, provided that one knows the relevant
terms in the string effective action.  In fact the nature of the 
possible higher-order terms is restricted severely by the fact that
they must satisfy certain universality conditions, and so although 
not much is known from direct string or sigma-model computations, 
it is possible to make natural conjectures for the structure of
such contributions.  This was discussed in detail in \cite{frposost},
where viable corrections at all orders in ${\a'}$ were proposed.
Using these terms, we analyse the associated corrections to the
various cohomogeneity one K\"ahler metrics enumerated above.

   The paper begins with a discussion of ${\a'}$ corrections in string
theory in Section \ref{revsec}.  We derive the explicit results for
corrections to six-dimensional cohomogeneity one metrics in Section
\ref{killspin}, and to eight-dimensional metrics in Section
\ref{cy6sec}.  In Section \ref{cy8sec} we derive results for the
contributions to the Euler numbers for the various Calabi-Yau
manifolds that come both from the volume integral of the Euler
integrand, and also from the boundary terms that must be included for
non-compact manifolds.  After the concluding Section \ref{concl} where
we comment on relations of our work to other $\alpha'$ correction
schemes found in the literature, we include appendices summarising
results by Chern on the structure of the boundary contributions to the
Euler number.

\section{$\a'$ Corrections in String Theory}\label{revsec}

   At leading order in string loops and $\a'$, the effective actions
in string theory coincide with type IIA, type IIB or type I
supergravities.  At higher order, these effective actions are
corrected by terms that involve higher derivatives, and  
higher powers of curvature and field strengths.

    Of particular interest are corrections in the type IIA and type
IIB string effective actions that are uncovered by studying
multi-particle graviton and graviton/dilaton scattering.  The leading
such corrections in graviton scattering, revealed by four-particle
amplitudes, imply the existence of terms in the effective action at
order ${\a'}^3$, associated with quartic invariants built from the
Riemann tensor.  The structure of these terms was discovered in early
papers on superstrings \cite{greenschwarz}, and a first analysis of
their implications for Calabi-Yau compactifications in string theory
was carried out in \cite{grosswitten}.  The results at first appeared
to exhibit puzzling discrepancies in relation to beta-function
calculations for sigma-model in K\"ahler background geometries
\cite{grisarua,grisarub}, but a closer study of the quartic-curvature 
terms from string theory showed that the two approaches were in agreement
\cite{freemanpope}.

     The superinvariant structure of the quartic curvature corrections
is closely related to that of the ultraviolet counterterms generally
anticipated at the three loop order in $D=4$ supergravity theories or
at corresponding lower orders in higher-dimensional theories. These
are known for minimal $N=1$, $D=4$ supergravity in component form
\cite{dks} where the full nonlinear structure can be written using
off-shell $N=1$ tensor calculus \cite{townsvanN} or in superspace
\cite{moura}. For the $N=8$ maximally extended theory (for which no
off-shell formalism exists), the structure is known at the quartic
order in fields \cite{kallosh,hst}. The $N=8$ quartic counterterm is
the dimensional reduction of the $D=11$ M-theory quartic correction,
which also corresponds to the type IIA string theory one-loop
correction \cite{afmn}. The full supersymmetric nonlinear structure of
the $D=11$ and $D=10$ quartic corrections is very complicated and
remains an unresolved issue, exacerbated by the absence of an
off-shell formalism for the maximally supersymmetric theories. A
Noether component-field program for supersymmetric construction of the
quartic invariants was launched in \cite{ggp,gsethi}. Beyond the
leading order, however, one has to begin to iteratively correct the
supersymmetry transformations as well; the current state of play for
this incomplete program is reviewed in \cite{pvhw}. A related issue is
the debate on the implications of the quartic corrections to the
structure of D-brane backgrounds in Refs \cite{skenderis,gstahn}. The
full component-field construction of the quartic corrections is quite
complicated, and we will not be concerned with the general case in the
present paper. Instead, we will concentrate on the structure of the
corrections as applied to K\"ahler manifolds without form-field
fluxes. This is a more tractable problem, and we shall see that it
sheds some light on the general construction.

    One of the outcomes of the analysis of ${\a'}^3$ corrections was
that a Ricci-flat K\"ahler Calabi-Yau metric that solves the internal
Einstein equations at leading order ceases to satisfy the equations
when the ${\a'}^3$ terms are present.  This was shown in beta-function
calculations in \cite{grisarua,grisarub}, and in string-scattering
calculations in \cite{freemanpope}.  The nature of the
corrections is relatively mild, in the sense that they imply a
distortion of the internal metric under which, for a suitable choice
of variables, it remains K\"ahler, but with the Ricci tensor deformed
away from zero in a manner that leaves its cohomology class unchanged.
It seems, therefore, that one can treat the corrections as
perturbations that smoothly deform the metric away from Ricci
flatness, provided that one considers a compactification whose scale
size is appreciably larger than the string length scale $\sqrt{\a'}$.

   In this paper, we shall focus principally on some explicit
calculations exploring the effect of the ${\a'}^3$ correction terms at
tree level in string theory.  It is useful, therefore, to begin by
summarising the detailed form of these terms. 

    In a four-point graviton scattering calculation performed at
string tree level in the light-cone gauge, one finds interactions
whose covariant description is provided by the contribution
\be
{\cal L} = -c\, {\a'}^3\, e^{-2\phi}\, Y_0\label{y0cont}
\ee
in the effective action, where $c$ is a constant,
\be
Y_0 \equiv \fft1{64}\, t^{i_1\cdots i_8}\, t^{j_1\cdots j_8}\, 
   R_{i_1 i_2 j_1 j_2}\cdots R_{i_7 i_8 j_7 j_8} \,,\label{y0def}
\ee
and the $t$-tensor is defined by
\be
t^{i_1\cdots i_8} \, M_{i_1 i_2}\dots M_{i_7 i_8} = 
   24 M_{i}{}^{j}\,  M_{j}{}^{k}\,M_{k}{}^{\ell}\,
M_{\ell}{}^{i} - 6 (M_{i}{}^{j}\, M_{j}{}^{i})^2
\ee
for an arbitrary antisymmetric tensor $M_{i_1 i_2}$.

   Further information about the quartic-curvature terms comes from 
considering dilaton/graviton scattering amplitudes, which imply that
the total contribution to the effective action must involve a quartic
curvature invariant that vanishes in Ricci-flat K\"ahler backgrounds.
This implies that, still in light-cone gauge, the contribution (\ref{y0cont})
is augmented to give
\be
{\cal L} = - c\, {\a'}^3\, e^{-2\phi}\, Y\label{ycont}
\ee
where
\be
Y \equiv \fft1{64}\, \td t^{i_1\cdots i_8}\, \td t^{j_1\cdots j_8}\, 
   R_{i_1 i_2 j_1 j_2}\cdots R_{i_7 i_8 j_7 j_8}= Y_0 + Y_1 + Y_2 
\label{ydefn}
\ee
and 
\be
\td t^{i_1\cdots i_8} = t^{i_1\cdots i_8} + \ft12 \ep^{i_1\cdots i_8}\,.
\ee
In (\ref{ydefn}) we are following the notation of \cite{frposost}, in
writing the contributions associated with 0, 1 and 2 $\epsilon$-tensor 
factors as $Y_0$, $Y_1$ and $Y_2$ respectively. 

   The expression $Y$ in (\ref{ydefn}) can be written in terms of a
path integral over $SO(8)$ fermion zero modes, in the form 
\cite{grosswitten}
\be
Y = \int d^8\psi_L\, d^8 \psi_R \exp(R_{ijk\ell}\, \bar\psi_L\, \Gamma^{ij}\,
\psi_L\, \bar\psi_R\, \Gamma^{k\ell}\, \psi_R)\,.
\ee
It was shown in \cite{freemanpope} that the variation of $Y$, 
specialised after variation to a Ricci-flat K\"ahler background, 
gives
\be
\fft{\delta Y}{\delta g^{ij}} = \nabla_{\hat i}\, \nabla_{\hat j}\, 
S_3\,,
\ee
where here, and in all subsequent formulae, we define 
\be
\nabla_{\hat i} \equiv J_i{}^j\, \nabla_j\,,
\ee
where $J_{ij}$ is the K\"ahler form, and  $S_3$ is given by
\be
S_3 = R_{abcd}\, R^{cdef}\, R_{ef}{}^{ab} - 2
      R_{acbd}\, R^{cedf}\, R_e{}^a{}_f{}^b\,.\label{s3exp} 
\ee

   The expression given in (\ref{ydefn}) does not immediately allow
itself to be re-expressed in a ten-dimensionally covariant fashion,
since it makes explicit use of the eight-index $\ep$-tensor of the
transverse eight-dimensional space in the light-cone gauge.  The
product of two $\ep$ tensors in $Y_2$ can be replaced by
antisymmetrised products of Kronecker deltas, thus allowing a
covariant extension to ten dimensions, but the term $Y_1$ linear in
$\ep$ admits no direct covariant extension.  This problem was studied in
\cite{frposost,grisaru2}, and a ten-dimensionally covariant Lagrangian was 
obtained.  Since it is important for our later purposes, we shall review
the construction of the ten-dimensionally covariant Lagrangian here, and
clarify some of the issues involved. 

   After straightforward combinatoric manipulations, one finds that
the term $Y_0$, defined in (\ref{y0def}), is a combination of quartic
Riemann-tensor invariants given by:
\be
Y_0= \ft3{16} (X_0 + 2X_1 + 16 X_2 -16 X_3 + 32 X_6 -8 X_7)\,.
\label{y0}
\ee
Here, we define the quartic Riemann tensor
invariants $X_0, \ldots, X_7$, as
\bea
X_0 &\equiv & (R_{abcd}\, R^{abcd})^2\,,\nn\\
X_1 &\equiv & R_{a_1 b_1 a_2 b_2}\, R^{a_2 b_2 a_3 b_3}\, 
R_{a_3 b_3 a_4 b_4}\, R^{a_4 b_4 a_1 b_1}\,,\nn\\
X_2 &\equiv& R_{a_1}{}^{a_2}{}_{b_1}{}^{b_2}\, 
          R_{a_2}{}^{a_3}{}_{b_2}{}^{b_3}\,
          R_{a_3}{}^{a_4}{}_{b_3}{}^{b_4}\,
          R_{a_4}{}^{a_1}{}_{b_4}{}^{b_1}\,,\nn\\
X_3 &\equiv & R_{a_1 a_2 b_1}{}^{b_2} R^{a_1 a_2}{}_{b_2}{}^{b_3}\,
             R_{a_3 a_4 b_3}{}^{b_4} \, R^{a_3 a_4}{}_{b_4}{}^{b_1}\,,\nn\\
X_4 &\equiv & R_{a_1 b_1 b_2 b_4}\, R^{a_1 b_3 a_4 b_4}\,  
              R^{a_2 b_1}{}_{a_3 b_3}\, R_{a_2}{}^{b_2 a_3}{}_{a_4}\,,\nn\\
X_5 &\equiv & R_{a_1 a_4 a_3 b_3}\, R^{a_2 b_2 a_3 b_3}\, 
R^{a_1}{}_{b_1 b_2 b_4}\, R_{a_2}{}^{b_1 a_4 b_4}\,,\nn\\
X_6 &\equiv & R_{a_1 a_2 b_1 b_2} \, R^{a_2 a_3 b_2 b_3}\,
R_{a_3 a_4 b_4}{}^{b_1} \, R^{a_4 a_1}{}_{b_3}{}^{b_4}\,,\nn\\
X_7 &\equiv & R_{a_1 a_2 b_1 b_2}\, R^{a_3 a_4 b_1 b_2}\, 
          R^{a_2}{}_{a_3 b_3 b_4}\, R_{a_4}{}^{a_1 b_3 b_4}\,.
\eea
Using the cyclic identity for the Riemann tensor, we have
\be
4X_4 + 4 X_5 - 4X_6 -X_7 = 0\,.
\ee

        The term $Y_2$ is proportional to the eight-dimensional Euler
integrand $E_8$, generalised to arbitrary dimension:
\be
Y_2 = 384\pi^4\, E_8\,,
\ee
with
\be
E_8 = \fft{105}{(4\pi)^4}\, R_{a_1 a_2}{}^{[a_1 a_2}\,
           R_{a_3 a_4}{}^{a_3 a_4}\, R_{a_5 a_6}{}^{a_5 a_6}\,
            R_{a_7 a_8}{}^{a_7 a_8]}\,.\label{e8exp}
\ee
We find that the exact expression for $Y_2$ is given by
\be
Y_2 = Y_2^\0 + Y_2^\1 + Y_2^\2\,,
\ee
where\footnote{Our grouping of terms in $Y_2$ is as follows.  $Y_2^\0$
denotes the terms involving only uncontracted Riemann tensors; $Y_2^\2$ 
denotes all terms involving at least one Ricci scalar; and $Y_2^\1$ 
denotes the remainder, namely terms without Ricci scalars and with at
least one Ricci tensor.}
\crampest
\bea
Y_2^\0\!\!\! &\equiv &\!\!\! \ft3{16} (X_0 + 2X_1 + 16 X_2 -16 X_3
-32 X_4 + 32 X_5)\,,\nn\\
Y_2^\1\!\!\! &\equiv&\!\!\! 
6R^{ab}\,(4 R_{a}{}^{cde}\, R_b{}^f{}_d{}^g\,
R_{efcg} + 2 R_a{}^d{}_b{}^e\, R_{echg}\, R_d{}^{chg} -
R_a{}^{cde}\, R_{bcfg}\, R_{de}{}^{fg})\nn\\
\!\!\!&&\!\!\!+6 R_{abef}\, R_{cd}{}^{ef}\, R^{ac}\, R^{bd} +
     12 R_{aebf}\, R_{c}{}^e{}_d{}^f\, 
(R^{ac}\, R^{bd} - R^{ab}\, R^{cd}) 
+ 12 R^{ac}\, R^b{}_c \, R_{adef}\, R_b{}^{def} \nn\\
\!\!\!&&\!\!\! -\ft32 R_{abcd}\, R^{abcd}\, R_{ef}\, 
R^{ef} -24 R_{abcd}\, R^{ac}\, 
R^{be}\, R^d{}_e - 6 R_{ab}\, R^{bc}\, 
R_{cd}\, R^{da} \,,\nn\\
Y_2^\2\!\!\! &\equiv &R (R_{abcd}\, R^{cdef}\,
R_{ef}{}^{ab} - 2
      R_{acbd}\, R^{cedf}\, R_e{}^a{}_f{}^b - R_{abcd}\, R^{abce}\, 
         R^d{}_e \nn\\
\!\!\!&&\!\!\! +\ft38 R\, R_{abcd}\, R^{abcd} + 6 R_{abcd}\,
R^{ac}\, R^{bd} 
     + 4 R_{ab}\, R^{bc}\, R_c{}^a - \ft32 R\, R_{ab}\, R^{ab} 
    + \ft1{16} R^3)\,.\label{y2}
\eea
\uncramp

   Of course the terms that are of quadratic or higher
order in the Ricci tensor or Ricci scalar are in any case irrelevant
here, since even after variation with respect to the metric,
their contributions will still vanish at order ${\a'}^3$ (since we can
impose the zero'th-order Ricci-flat K\"ahler background equations
on these corrections that carry an explicit ${\a'}^3$ factor, after
varying to derive the equations of motion).
However, terms linear in the Ricci tensor or Ricci scalar {\it will}
contribute to the equations of motion at this order, since the
variations of the Ricci terms will themselves give non-vanishing
contributions.  Thus we just need
\bea
Y_2^\1  & = &
6R^{ab}\,(4 R_{a}{}^{cde}\, R_b{}^f{}_d{}^g\,
R_{efcg} + 2 R_a{}^d{}_b{}^e\, R_{echg}\, R_d{}^{chg} -
R_a{}^{cde}\, R_{bcfg}\, R_{de}{}^{fg}) +\cdots \,, \nn\\
Y_2^\2  & = & R (R_{abcd}\, R^{cdef}\,
R_{ef}{}^{ab} - 2
      R_{acbd}\, R^{cedf}\, R_e{}^a{}_f{}^b) + \cdots\,,\label{ylins}
\eea
where the ellipses represent terms of quadratic or higher order in the
Ricci tensor or scalar, which can be neglected in the present
discussion.

 Some comment is appropriate on why it is convenient to retain at
least some terms involving the Ricci tensor and Ricci scalar. Such
terms can of course be adjusted by field redefinitions that cause
terms proportional to the leading $(\alpha')^0$ field equations to mix
with the quartic corrections. One could decide to use such field
redefinitions to eliminate the Ricci tensor and Ricci scalar terms
retained in (\ref{ylins}); one could also proceed further and decide
to extract all the Ricci terms from the curvatures in $Y_2^\0$,
retaining only Weyl tensors in the place of the curvatures. Different
choices of this sort correspond to different choices of field
renormalisations. Our choice to retain the terms in (\ref{ylins}) will
have the virtue of allowing us to preserve the K\"ahler structure of
the internal manifold. We will return to the matter of field
redefinitions in the Conclusion, in which we compare our results to
those of \cite{tseytlin}.

   Another related issue is that of renormalisation-scheme dependence
of sigma-model beta functions. As discussed in \cite{grisaru2}, the
coefficients of terms in $L$ that are linear in the Ricci {\it scalar}
(these reside in $Y_2^\2$ in our description) are not determined
within sigma-model beta function calculations, since they produce
contributions to the equations of motion that are absorbable by
redefinitions of the sigma-model scalar fields.  To see this more
clearly, recall that what one calculates directly from the sigma model
are the renormalisation group beta functions, which are taken to give
effective field equations for the massless modes when set to zero. The
effect of an actively-viewed general coordinate transformation with
parameter $V_i$ on the metric is $\delta g_{ij}=
\nabla_{(i}V_{j)}$. Moreover, terms in the variation of the effective
action proportional to $g_{ij}$ should correspond to the dilaton beta
function \cite{cfmp,ckp}. Thus, since the variation of $\sqrt{-g}RW$
gives contributions to the gravitational equation of the form
$\nabla_i\nabla_jW-g_{ij}\nabla^2W$ plus terms containing $R_{ij}$,
these contributions can be absorbed into coordinate-transformations of
the metric and dilaton. Such terms are {\em scheme-dependent} from the
sigma-model point of view, and can be changed by changes of
regularisation and subtraction procedure.  Nonetheless, having chosen
a specific renormalisation scheme, the coefficients of terms linear in
the Ricci scalar do have significance. By contrast, the terms linear
in the Ricci {\it tensor} (residing in $Y_2^\1$ in our description)
are not subject to these scheme-dependent ambiguities.

      Let us now look at the $Y_1$ term, which does not admit an
obvious generalisation to a fully ten-dimensionally covariant
expression whilst maintaining all of the necessary features that its
exhibits in special backgrounds.  This issue was explored in detail in
\cite{frposost}, where it was noted that by a topological property of
K\"ahler manifolds the integral of $Y_1$ could be replaced by the
integral of $-2Y_2^\2$.  In this paper, we observe that in an
eight-dimensional Ricci-flat K\"ahler background, $Y_1$ can in fact be
be directly expressed as
\be
Y_1=-2Y_2^\0\,.\label{y1y0res}
\ee
This can be seen by noting that, viewed as 8-forms, we have
\be
Y_1 = 6 \tr\Theta^4 - \ft32 (\tr\Theta^2)^2\,,\qquad
Y_2 = \ft1{16} \ep_{i_1\cdots i_8}\, \Theta^{i_1 i_2}\cdots
\Theta^{i_7 i_8}\,.
\ee
where $\Theta_{ab}=\ft12 R_{abmn}\, dx^m\wedge dx^n$.
Now in a K\"ahler metric, with K\"ahler form $J_{ij}$, we have
$\ep_{i_1\cdots i_8}= 105 J_{[i_1 i_2}\cdots J_{i_7 i_8]}$.  After
straightforward combinatoric manipulations, we find that after
substituting this into the expression for $Y_2$, and using $J_{ac}
J_{bd}\, \Theta^{cd}= \Theta_{ab}$, then the terms in $Y_2$ where
there is no contraction of the form $J_{ab}\, \Theta^{ab}$ (\ie the
terms where there is no contraction of the Riemann tensor to give a
Ricci tensor) are given by
\be
-3  \tr\Theta^4 + \ft34 (\tr\Theta^2)^2\,.
\ee
In other words, we have the result that (\ref{y1y0res}) holds 
in an eight-dimensional Ricci-flat K\"ahler background.  (The effect of
$Y_1$ in correcting special-holonomy backgrounds with flux in M-theory was 
considered in \cite{cvetic}.)

   Based on the topological argument mentioned above, it was therefore 
conjectured in \cite{frposost} that the appropriate
ten-dimensionally covariant generalisation of the light-cone
Lagrangian (\ref{ydefn}) at $\alpha'^3$ order should be given
by
\be
{\cal L} = - c\, \alpha'^3\, e^{-2\phi}\, 
(Y_0 - Y_2)\,.\label{lans}
\ee
Our observation in Eqn (\ref{y1y0res}) lends further support to this
proposal.  However, it is really only by performing a variation of
(\ref{lans}) explicitly that one can give a complete verification, since
(\ref{lans}) was obtained by the potentially
hazardous procedure of substituting the Ricci-flat K\"ahler background
condition into the Lagrangian, prior to its variation.

   The full Lagrangian, up to this order, should take the form
\be
{\cal L} = \sqrt{-g}\, e^{-2\phi}\, (R +4 (\del\phi)^2 - c\, {\a'}^3
\, Q)\,,\label{qlag}
\ee
where $Q$ is a ten-dimensionally covariant function whose variation 
$Q_{ij}\equiv \delta Q/\delta g^{ij}$
gives
\be
Q_{ij}
= \nabla_{\hat i}\, \nabla_{\hat j}\, S_\3\,,
\ee
when specialised to a Ricci-flat K\"ahler background.  (We are allowed
to employ the leading-order Ricci-flat K\"ahler background equations here,
{\it after} the variation, since there is already an explicit ${\a'}^3$ 
factor in the term involving $Q$.)  The dilaton and Einstein equations 
following from (\ref{qlag}) give
\bea
&&R + 4\, \square \phi - 4(\del\phi)^2 - c\, {\a'}^3\, Q =0\,,\nn\\
&&R_{ij} + 2\nabla_i\nabla_j\, \phi - c\, {\a'}^3\, Q_{ij}  -
\ft12 (R + 4\, \square \phi - 4(\del\phi)^2 - c\, {\a'}^3\, Q) \, 
g_{ij} = 0\,.
\eea
Specialising to a Ricci-flat K\"ahler background and substituting the
former into the latter equation in the ${\a'}^3$ terms as discussed
above, gives
\be
R_{ij} = \nabla_{\hat i}\nabla_{\hat j}\, S_3 - 2 \nabla_i\nabla_j\, \phi
\label{tlein}
\ee
Taking the trace of this, substituting into the dilaton equation and
neglecting the term $(\del\phi)^2$, since it would be of order
${\a'}^6$, gives 
\be
\square(2\phi + c\, {\a'}^3\, S_3)=0\ ,\label{dileqn}
\ee 
so we can take \cite{cfpss}
\be
\phi= -\ft12 c\, {\a'}^3\, S_3\ .\label{dilcorr}
\ee 
Finally, (\ref{tlein}) then implies that we have 
\be
R_{ij} = c\, {\a'}^3\, (\nabla_i\nabla_j + 
                   \nabla_{\hat i} \nabla_{\hat j})\, S_3\,.\label{rij}
\ee
Since this is the desired result, it therefore remains to establish
that indeed we can take $Q$ to be given by 
\be
Q=Y_0-Y_2\,,\label{qdef}
\ee
as proposed in \cite{frposost} and in agreement with (\ref{lans}).

    As was noted in \cite{frposost}, and as is evident from (\ref{y0})
and (\ref{y2}), the Riemann tensor structure appearing in the
effective action (\ref{lans}) is much simpler than that found in each
individual term in the $Y$'s; the full expression in (\ref{lans})
is given by
\be
{\cal L} = e^{-2\phi}\,\sqrt {-g}\, \Big[ R + 4(\del\phi)^2 -c\, {\a'}^3\,  
( 12(X_6-X_5) - Y_2^\1 - Y_2^\2)\Big]\,.
\label{ycomplete}
\ee

    A convenient way to establish that (\ref{lans}) gives the desired
form of the equation of motion (\ref{rij}) is first to address a
slightly different problem, in which one considers the beta function
for a pure $N=2$ supersymmetric sigma model without a dilaton.  In
this case without a dilaton in the model, the vanishing of the beta
function at the four-loop level gives rise once again to the condition
(\ref{rij}).  One can ask whether there exists an action for this
beta-function equation, and if so, how it relates to the desired
string-theory action discussed above.  Let us write the beta-function
Lagrangian as
\be
{\cal L}_\sigma = \sqrt{g}\, (R - c\, {\a'}^3\, P)\,.\label{lagans}
\ee
A natural ansatz for $P$ is to take
\be
P= Y_0 - Y_2^\0 + c_1\, Y_2^\1 + 
            c_2 \, Y_2^\2\,,\label{pans}
\ee
where $c_1$ and $c_2$ are constants to be determined.  (By contrast,
the coefficient of $Y_2^\0$ is determined by the requirement that $P$
should vanish in a Ricci-flat K\"ahler background.)  As we have
already mentioned, $c_1$ and $c_2$ can be adjusted by field
redefinitions. Nonetheless, if one wants the specific
K\"ahler-preserving structure of the corrected Einstein equation given
in (\ref{rij}), then $c_1$ and $c_2$ are determined uniquely.

This problem of finding an action that produces the sigma-model beta
function as its equation of motion was studied in \cite{grisaru2}.
Here, we shall not perform an explicit variation of (\ref{lagans}),
but rather we shall make use of special cases of cohomogeneity one
metrics that admit Ricci-flat K\"ahler solutions in order to determine
the coefficients $c_1$ and $c_2$ in (\ref{pans}) by requiring
consistency with (\ref{rij}).  We can do this by simply substituting
the general cohomogeneity one metric into (\ref{lagans}) and then
obtaining equations of motion by varying the metric
functions.\footnote{This is a valid procedure provided that one
substitutes the most general form of metric invariant under the
isometries of the homogeneous level surfaces.  Such a shortcut to
obtaining a consistent truncation has also been employed in Ref.\
\cite{skenderis}.}  The calculations must be performed for
eight-dimensional K\"ahler metrics in order to pin down fully the
structure of the Lagrangian.  In practice, the calculations are of a
sufficient degree of complexity that a computer is helpful.

   We have carried out this procedure for many of the metric examples
discussed in the later sections of the paper, and we find universal
results for the two coefficients $c_1$ and $c_2$, namely $c_1=-1$,
$c_2=-2$.  Thus we conclude that the Lagrangian (\ref{lagans}), 
with
\be
P= Y_0 - Y_2^\0  - Y_2^\1 - 
           2Y_2^\2 \,,\label{pans2}
\ee
gives rise to the $N=2$ sigma-model beta function.\footnote{In 
\cite{grisaru2}, the Lagrangian 
${\cal L}_{\rm gz} = \sqrt{g}\, [R - c\, 
\alpha'^3\, (Y_0 - Y_2^\0 - Y_2^\1 -\ft43 Y_2^\2) ]$ is obtained.
Evidently, therefore, the scheme employed in \cite{grisaru2} differs from
ours, for which the coefficients $c_1$ and $c_2$ in (\ref{lans}) are 
uniquely defined by the fact that in a K\"ahler 
background we have (\ref{rij}).}  In particular, note that
the variation of (\ref{lagans}) gives 
\be
R_{ij} - \ft12 R\, g_{ij} - c\, {\a'}^3\, P_{ij}=0\,,
\ee
where $P_{ij} =\delta P/\delta g^{ij}$, and we have used the fact that
$P$ itself vanishes in the Ricci-flat K\"ahler background.  Since
$R_{ij}= c\, {\a'}^3\, ( \nabla_i\nabla_j 
+ \nabla_{\hat i}\nabla_{\hat j})\,  S_3$, it follows that 
\be
P_{ij} = \nabla_i\nabla_j \, S_3 + \nabla_{\hat i}\nabla_{\hat j}\, 
S_3 - \square S_3\, g_{ij}\,.\label{pvar}
\ee

   Having determined the variation of $P$ in (\ref{pans2}), we can now
go back to the tree-level string effective Lagrangian (\ref{qlag})
including the dilaton, where $Q$ is given by (\ref{qdef}).  Comparing
(\ref{qdef}) and (\ref{pans2}), we see that
\be
Q = P + Y_2^\2\,.
\ee
The relevant terms in $Y_2^\2$ (\ie those linear in $R$) are given by
$Y_2^\2 = R\, S_3 + \cdots$.  Using
\be
\delta R = (R_{ij} - \nabla_i\nabla_j + g_{ij}\, \square)\, \delta g^{ij}\,,
\ee
it follows that the variation of $Y_2^\2$, in a Ricci-flat K\"ahler
background, gives
\be
\fft{\delta Y_2^\2}{\delta g^{ij}} = - \nabla_i\nabla_j \, S_3
 + \square S_3\, g_{ij}\,.
\ee
Hence, from (\ref{pvar}), it follows that
\be
Q_{ij} = \nabla_{\hat i}\nabla_{\hat j}\, S_3\,.\label{qvar}
\ee
Thus we have verified that taking $Q$ to be given by (\ref{qdef}) does
indeed give the correct string effective Lagrangian.

   Finally, we shall make a remark about the structure of the terms
proportional to $g_{ij}$ coming from the variation of the Lagrangians
we have been considering.  These terms are of significance because
they should be absent in the metric beta function for the $N=2$ sigma
model.  The calculation is slightly subtle, since not only do such
terms arise from the obvious source $\delta \sqrt{g}/\delta g^{ij} =
-\ft12 \sqrt{g}\, g_{ij}$, but also from the variation of metrics in
$R_{ij}$ and $R$ in $Y_2^\1$ and $Y_2^\2$ respectively.  In fact one 
finds
\be \fft{\delta Y_2^\1}{\delta g^{ij}} = \ft12 \nabla_k\nabla_\ell\,
S^{k\ell}\, g_{ij} + \cdots\,,\qquad 
\fft{\delta Y_2^\2}{\delta g^{ij}} = \square\, S_3\, g_{ij} + \cdots\,,
\ee
where the ellipses represent the terms not proportional to $g_{ij}$, 
and $S_{ij}$ is defined by $Y_2^\1= R^{ij}\, S_{ij}$ (see (\ref{ylins})).
One can show that in a Ricci-flat background $\nabla_i\nabla_j\, S^{ij} =
-2\, \square\, S_3$, and hence this explains the $g_{ij}$ term in $\delta P/
\delta g^{ij}$ in (\ref{pvar}), and the absence of the $g_{ij}$ terms
in $\delta Q/\delta g^{ij}$ in (\ref{qvar}).

\section{Killing Spinors, Integrability
Conditions and Field Equations}\label{killspin}

   As we discussed in Section \ref{revsec}, the equations of motion
for the internal Calabi-Yau manifold $K_6$ in a (Minkowski)$_4\times
K_6$ solution in string theory receive non-vanishing corrections at
orders ${\a'}^3$ and above.  These are of the form
\be
R_{ab} = (\nabla_a\nabla_b + \nabla_{\hat a}\nabla_{\hat b})\, S
\,,\label{ricmod}
\ee
where as usual $\nabla_{\hat a}\equiv J_a{}^b\, \nabla_b$, 
\be
S= \sum_{n=3}^\infty {\a'}^n\, S_n\,,\label{ssum}
\ee
and from this point onwards, we shall choose units where constant $c$
appearing in (\ref{y0cont}) and subsequent formulae in Section
\ref{revsec} is set to unity.  Here $S_n$ are certain invariants built
from products of $n$ Riemann tensors.  Multiplying (\ref{ricmod}) by
$J_k{}^j$, we can recast it in terms of differential forms as
\be
\varrho = d\hat d\, S\,,\label{ricmod2}
\ee
where the Ricci-form $\varrho$ is defined by 
\be
\varrho_{ab}\equiv \ft12 R_{abcd}\, J^{cd} = J_b{}^c\, R_{ac}\,,
\ee
and $\hat d f\equiv \del_{\hat a}f\, e^a$.  Note that we therefore have
\be
d=\del+\bar\del\,,\qquad  \hat d = -\im\, (\del -\bar\del)\,,
\ee
where $\del$ and $\bar\del$ are the holomorphic and anti-holomorphic 
exterior derivative operators.  Thus (\ref{ricmod2}) is equivalent to
$\varrho= 2\im\, \del\bar\del S$, showing that the right-hand 
side can be viewed
as a cohomologically trivial $(1,1)$ deformation of the leading-order
Ricci-flat condition. 

    Equations (\ref{ricmod}) or (\ref{ricmod2}) define a
deformation from Ricci-flatness in which K\"ahlerity is preserved.
In fact the solution will also continue to be
supersymmetric.  It was shown in \cite{cfpss} that a K\"ahler metric
satisfying (\ref{ricmod}) admits Killing spinors that satisfy the modified
equation
\be
\nabla_a\eta  + \ft{\im}{2}\, (\del_{\hat a}S)\,\eta=0\,,\label{kseqn}
\ee
where $\nabla\eta=d\eta +\ft14 \omega_{ab}\, \Gamma^{ab}\, \eta$.
In fact (\ref{kseqn}) can be written as
\be
\nabla\eta + \ft{\im}{2}\, (\hat d S)\, \eta=0\,,
\ee
whose integrability condition $(\nabla + \ft{\im}{2}\, (\hat d S))^2\eta=0$
is
\be
\ft14 \Theta_{ab}\, \Gamma^{ab}\, \eta +\ft{\im}{2}\, \varrho\, \eta=0\,.
\ee
Writing this in components, $\ft14 R_{abcd}\, \Gamma^{cd}\, \eta
+ \ft{\im}{2}\, \varrho_{ab}\, \eta=0$, and multiplying by
$\Gamma^c$, it is manifest that the integrability condition is
satisfied, by virtue of the holomorphicity condition 
\be
\Gamma_a\, \eta = -\im\, \Gamma_{\hat a}\, \eta\,.\label{holomorphicity}
\ee

    We shall make use of these observations about the existence of
Killing spinors in the following subsections, where we study the
effect of the right-hand side of (\ref{ricmod}) in deforming
previously-known complete Ricci-flat K\"ahler metrics.  Specifically,
from the existence of the Killing spinors we shall be able to derive
first-order systems of differential equations for the perturbed
metrics, and hence to construct explicit solutions at order 
${\a'}^3$.  We shall apply the technique to three types of six-dimensional
Ricci-flat K\"ahler starting points, namely the resolved conifold, the
deformed conifold, and the $\R^2$ bundles over $S^2\times S^2$ or
$\CP^2$.

    When we construct fully explicit perturbative solutions, we shall
focus first on the term $S_3$ in (\ref{ssum}), corresponding to order
${\a'}^3$.  This is the cubic curvature invariant, given in
(\ref{s3exp}), that arises in the type IIA and IIB string theories.
We shall also consider corrections at higher order in $\a'$, namely
${\a'}^4$ and ${\a'}^5$.  Candidate terms at these, and all higher
orders, that satisfy the highly-restrictive {\it universality
conditions} were conjectured in \cite{frposost}.\footnote{Once one
considers corrections beyond $O({\a'}^5)$ the discussion becomes 
considerably more complicated, because now one can no longer simply
impose the zero'th-order Ricci-flat K\"ahler background equations 
on the variations of the correction terms in the Lagrangian.  This 
is because the curvature of the true solution itself has $O({\a'}^3)$
deviations from its zero'th-order form, and these deviations therefore
make $O({\a'}^6)$ contributions to the variations of the 
corrections that carry explicit factors of ${\a'}^3$ and above.} 
 
  The universality conditions
arise from the fact that, in a sigma-model beta-function calculation,
since K\"ahler or hyper-K\"ahler target-space background are but
specialisations of generic Riemannian backgrounds, it follows that the
known special forms of the beta-functions in K\"ahler or
hyper-K\"ahler backgrounds must be expressible in terms of purely
Riemannian quantities.  Thus, specifically, the known form of the beta
function in a Ricci-flat K\"ahler background, $\beta_{ab} \sim
(\nabla_a \nabla_b + J_a{}^c\, J_b{}^d\, \nabla_c\nabla_d)\, S$, must
be expressible in purely Riemannian terms, \ie without the use of the
complex structure $J_a{}^b$.  Similarly, since the beta-function is
known to vanish in hyper-K\"ahler backgrounds, the Riemannian
expression must have the property of vanishing under this
specialisation.

   The universality conditions apply similarly to the $\alpha'$
corrected Killing spinor conditions, since these should ultimately
have an origin in vanishing gravitino conditions
$\delta\psi_m^\alpha=0$. Indeed, partial results for the corrected
gravitino transformation have been derived in $D=11$ and $D=10$
supergravities via a Noether supersymmetrisation procedure for the
quartic corrections to the action \cite{pvhw}. Conversely, one can use
the universality conditions as a guide to finding the structures of
correction terms. We observe that the universality properties of $S$
allow the corrected Killing spinor condition (\ref{kseqn}) to be
written without the use of complex structures.

   There are in fact two different such forms, equivalent when
evaluated on Ricci-flat K\"ahler spaces: one with a $\Gamma_{mnpqrs}$
structure \cite{cfpss} and one with a $\Gamma_{mn}$ structure
\cite{frposost}. The six-$\Gamma$ form is
\be
\nabla_i\eta  - \ft34
\nabla_sR_{irkl}R^s{}_{tmn}R^{tr}{}_{pq}\Gamma^{klmnpq}\eta =
0\,,\label{6gamma}
\ee
plus terms that vanish for the
leading-order Ricci-flat K\"ahler solution. The two-$\Gamma$ form is
\be
\nabla_i\eta  - 6\nabla_sR_{ipkl}R^{stln}R_t{}^p{}_n{}^c\Gamma_{ck}\eta = 0\,.
\label{2gamma}
\ee
The equivalence of the
two forms for Ricci-flat K\"ahler spaces is established by dualising
$\Gamma_{i_1\ldots i_6} = -\ft12 \epsilon_{i_1\ldots
i_6jk}\Gamma_9\Gamma^{jk}$, picking $\Gamma_9\eta=\eta$ and using
$J_{[ij}J_{kl}J_{mn}J_{pq]}=\ft1{105}\epsilon_{ijklmnpq}$ and
$(\Gamma_{ij}+\Gamma_{\hat i\hat\j})\eta=2\im J_{ij}$, which follows
from the Killing spinor holomorphicity condition
(\ref{holomorphicity}). Using the hat-flipping rules to eliminate the
complex structures and dropping Ricci tensor terms, one obtains the
equivalence of the two forms (\ref{6gamma},\ref{2gamma}). This
equivalence for Ricci-flat K\"ahler spaces illustrates that the full
$D=10$ or $D=11$ expression could be a mixture of various forms that
become equivalent when evaluated on Ricci-flat K\"ahler spaces; this
impression is borne out by the partial results in \cite{pvhw}.

 From a geometrical point of view, the two-$\Gamma$ form
(\ref{2gamma}) is noteworthy because it shows that the $\alpha'$
corrections can be viewed as requiring a connection with {\em torsion}
in the Killing spinor connection, with respect to which one simply has
$\nabla^{\rm corr}_i\eta=0$. In order to preserve the K\"ahler
hat-flipping rule $R_{ijk\hat l}=-R_{ij\hat k l}$, this corrected
connection with torsion must remain hermitean, $\nabla^{\rm
corr}_iJ_{jk}=0$.

In the next sections, we will use the corrected Killing equation
(\ref{kseqn}) to work out explicitly the changes to a set of
non-compact Calabi-Yau manifolds.

\section{Explicit Non-compact Calabi-Yau Examples in $D=6$}\label{cy6sec}
\subsection{Corrections to the resolved conifold}\label{resconsec}

    To describe the metric on the resolved and deformed conifolds, it
is convenient to introduce the left-invariant 1-forms $\sigma_i$ and 
$\Sigma_i$ for two copies of $SU(2)$.  These satisfy
\be
d\sigma_i = -\ft12 \ep_{ijk}\, \sigma_j\wedge \sigma_k\,,\qquad
d\Sigma_i = -\ft12 \ep_{ijk}\, \Sigma_j\wedge \Sigma_k\,.
\ee
We write the metric on the resolved conifold as
\be
ds_6^2 = dt^2 + a^2\, (\Sigma_1^2 + \Sigma_2^2) + b^2\, (
\sigma_1^2 + \sigma_2^2) + c^2\, (\Sigma_3-\sigma_3)^2\,,\label{rescon}
\ee
and choose the natural vielbein basis
\be
e^0=dt\,,\quad e^1=a\, \Sigma_1\,,\quad e^2=a\, \Sigma_2\,,\quad
e^3=b\, \sigma_1\,,\quad e^4=b\, \sigma_2\,,\quad e^5=c\, 
(\Sigma_3-\sigma_3)\,,
\ee
where $a$, $b$ and $c$ are functions of $t$. The principal orbits are
$T^{1,1}=(S^3\times S^3)/U(1)$, the denominator corresponding to
the diagonal $U(1)$ with left-invariant 1-form $(\Sigma_3+\sigma_3)$.

   The torsion-free spin connection is easily calculated.  It is
convenient to present it by giving the Lorentz-covariant exterior
derivative $\nabla=d+\ft14 \omega_{ab}\, \Gamma^{ab}$ that acts on
spinors, with vielbein components $\nabla_a$ defined by $\nabla=e^a\,
\nabla_a$:
\bea
&&\nabla_0 = d_0\,,\nn\\
&&\nabla_1 = 
  d_1 - \fft{\dot a}{2a}\, \Gamma_{01} - \fft{c}{4a^2}\, \Gamma_{25}\,,
\qquad
\nabla_2 = 
  d_2 - \fft{\dot a}{2a}\, \Gamma_{02} + \fft{c}{4a^2}\, \Gamma_{15}\,,\nn\\
&&\nabla_3 = 
  d_3 - \fft{\dot b}{2b}\, \Gamma_{03} + \fft{c}{4b^2}\, \Gamma_{45}\,,
\qquad
\nabla_4 = 
  d_4 - \fft{\dot b}{2b}\, \Gamma_{04} - \fft{c}{4b^2}\, \Gamma_{35}\,,\nn\\
&&\nabla_5 = d_5 - \fft{\dot c}{2c}\, \Gamma_{05} +
   \fft{c^2-a^2}{4a^2\, c}\, \Gamma_{12} + \fft{b^2-c^2}{4 b^2\, c}\, 
              \Gamma_{34}\,.\label{resspincon}
\eea
(There are also additional terms $\omega_{12}^{\rm extra}= 
\omega_{34}^{\rm extra}= -\ft12 (\Sigma_3+\sigma_3)$ which lie outside the
$S^3\times S^3)/U(1)$ coset.  These project out in the coset construction.
See \cite{ricciflat} for a further discussion.)

   After calculating the curvature from the spin connection, one finds that
the Ricci tensor is given by
\bea
R_{00} &=& -\fft{2\ddot a}{a} -\fft{2\ddot b}{b} -\fft{\dot c}{c}
\,,\nn\\
R_{11} &=& R_{22} = -\fft{\ddot a}{a} - \fft{\dot a^2}{a^2} -
\fft{2\dot a\,\dot b}{a\,b} -\fft{\dot a\,\dot c}{a\,c} +
\fft{1}{a^2} -\fft{c^2}{2a^4}\,,\nn\\
R_{33} &=& R_{44} = -\fft{\ddot b}{b} -\fft{\dot b^2}{b^2} -
\fft{\dot b\,\dot c}{b\,c}-\fft{2\dot a\, \dot b}{a\,b}
+\fft{1}{b^2} -\fft{c^2}{2b^4} \,,\nn\\
R_{55} &=& -\fft{\ddot c}{c} - \fft{2\dot a\, \dot c}{a\,c} -
\fft{2\dot b\, \dot c}{b\,c} + \fft{c^2}{2a^4} + \fft{c^2}{2b^4}
\,,\label{resricci}
\eea
The corrected equations of motion (\ref{ricmod}) 
for the system are therefore given by
\be
R_{00}=R_{55}= \ddot S + \fft{\dot c}{c}\, \dot S\,,\qquad
R_{11}=R_{22}= \fft{2 \dot a}{a}\, \dot S\,,\qquad
R_{33}=R_{44}= \fft{2 \dot b}{b}\, \dot S\,,\label{xxyy}
\ee
where the Ricci tensor is given by (\ref{resricci}).

   The system of first-order equations that govern the Ricci-flat
resolved conifold itself can easily be derived from
(\ref{resspincon}), by requiring the existence of a
covariantly-constant spinor, satisfying $\nabla_a\eta=0$.  We can see
by inspection that a spinor with constant components, and satisfying
the projection conditions
\be
\Gamma_{05}\eta= \Gamma_{12}\eta = -\Gamma_{34}\eta = \im\, \eta
\ee
will be covariantly-constant provided that the first-order equations
\be
\dot a = -\fft{c}{2a}\,,\qquad \dot b = -\fft{c}{2b} \,,\qquad
\dot c = -1 + \fft{c^2}{2a^2} + \fft{c^2}{2b^2}\label{resfo}
\ee
hold.  We can also see that if $\eta$ is normalised so that $\bar\eta\eta=1$,
then the relation $J_{ab}= -\im\,\bar\eta\, \Gamma_{ab}\, \eta$ gives the 
K\"ahler form,
\be
J= e^0\wedge e^5 + e^1\wedge e^2 - e^3 \wedge e^4\,.
\ee

   It is evident from (\ref{kseqn}) that if we now turn on the right-hand
side in (\ref{ricmod}), the previous Killing-spinor equations will 
receive a modification only in the ``5'' direction, \ie
\bea
\nabla_a\eta &=& 0\,,\qquad 0\le a\le 4\,,\nn\\
\nabla_5\eta -\ft{\im}{2}\, \dot S\, \eta &=&0\,.
\eea
We can immediately see, therefore, that the previous first-order equations
for the Ricci-flat case, given in (\ref{resfo}), will be modified to
become
\be
\dot a = -\fft{c}{2a}\,,\qquad \dot b = -\fft{c}{2b} \,,\qquad
\dot c = -1 + \fft{c^2}{2a^2} + \fft{c^2}{2b^2} -c\, \dot S\,.
\label{resfomod}
\ee
It should be emphasised that these are exact equations, valid for any 
function $S(t)$.  In other words, for any $S$ the first-order equations
(\ref{resfomod}) imply that that the metric (\ref{rescon}) 
will satisfy the modified Einstein equations (\ref{xxyy}).  Note that
analogous first-order equations were obtained by a different method, and
in a different scheme, in \cite{tseytlin}.

   To solve the modified first-order equations, it is convenient 
to introduce a new radial coordinate $\rho$,
defined by $dt= - c^{-1}\, d\rho$.  The first-order equations
(\ref{resfomod}) become
\be
a'=\fft{1}{2a}\,,\qquad b'=\fft{1}{2b}\,,\qquad 
c' = \fft1{c} - \fft{c}{2a^2} -\fft{c}{2b^2} - c\, S'\,,
\label{primefo}
\ee
where a prime denotes a derivative with respect to $\rho$. The functions
$a$ and $b$ can be easily solved, giving
\be
a^2=\rho + \ell_1^2 \,,\qquad b^2 = \rho + \ell_2^2\,,\label{abressol}
\ee
and solving for $c$ we find
\be
c^2 = \fft{2}{a^2\, b^2}\, e^{-2S}\, 
\int_0^\rho a(x)^2\, b(x)^2\, e^{2S(x)} \, dx\,.\label{cdifeq}
\ee
If $S$ were an externally-specified source term, then this would represent an
exact solution to the modified Einstein equations (\ref{ricmod}).  
It should, however, be emphasised that in the present paper we are
taking $S$ to be given by the higher-order corrections to the string 
effective action, and so $S$ itself is a function of the curvature, and 
hence a function of $a$, $b$, $c$ and their derivatives.  In this context,
therefore, (\ref{cdifeq}) is an integro-differential equation, which in
principle determines $c$. 

   We can give an explicit solution by linearising the system.  Thus
we send $S\longrightarrow \varepsilon\, S$, write
\be
c = \bar c\, (1 + \varepsilon \, f)\,,\label{cfexp}
\ee
and now work only to first order in $\varepsilon$.  (Note that since
the $a$ and $b$ equations in (\ref{primefo}) do not involve $S$, their
solutions, given in (\ref{abressol}), remain unchanged by the
perturbation.)  Substituting (\ref{cfexp}) into (\ref{primefo}), we
find that $f$ can be solved explicitly, to give
\be
f = \fft{2 \bar P}{a^2\, b^2\, \bar c^2} - \bar S\,.\label{fsolution}
\ee
Here $\bar S$ denotes the curvature invariant appearing in (\ref{ssum}),
evaluated in the unperturbed Ricci-flat metric (\ie in terms of $a$ and $b$
and the unperturbed metric function $\bar c$).  The function
$\bar P$ is defined by
\be
\bar P(\rho) = \int_0^\rho a(x)^2\, b(x)^2\, \bar S(x)\, dx\,.
\ee
 
    If we consider the specific example of the $n=3$ term in (\ref{ssum}),
we may note that, being the Euler integrand in six dimensions (modulo
Ricci tensor terms that vanish in the background), $\sqrt{g}\, S_3$
given in (\ref{s3exp}) is expressible (locally) as a total derivative.  
In the coordinate gauge we are using here, we therefore have
\be
\bar S_3 = \fft{1}{a^2\, b^2}\, \fft{d \bar P}{d\rho}\,.\label{barS3}
\ee
An algebraic computer calculation shows that $\bar P$ is given by
\be
\bar P=k+ \fft{ 18 a^2\, b^2\, (a^2+ b^2)\, \bar c^2 
    + 12(a^2 + b^2)\, \bar c^6 - 3(3 a^2 + b^2)(3
b^2 + a^2)\, \bar c^4}{ a^4\, b^4}\,,\label{pres}
\ee
where $k$ is an arbitrary constant. 

  The Ricci-flat resolved conifold solution \cite{conifold} is given
by setting $\ell_1=0$ in (\ref{abressol}), and evaluating $\bar c$ by
setting $S=0$ in (\ref{cdifeq}).  There is an $S^2$ bolt at $\rho=0$,
and the metric approaches the cone over $T^{1,1}$ at large $\rho$.  We
have
\be
a^2 = \rho\,,\qquad b^2 = \rho + \ell^2\,,\qquad \bar c^2 = 
\fft{\rho\, (2\rho + 3\ell^2)}{3(\rho+\ell^2)}\,,\label{barsolres}
\ee
where we have replaced $\ell_2$ by $\ell$.
We obtain a regular solution for $f$, which remains finite for $0\le \rho
\le\infty$, by choosing $k=-9$ in (\ref{pres}).  This gives
\bea
\bar P &=& \fft{\rho^2\, (\rho+2\ell^2)(7\rho^2 + 21\rho\, \ell^2 
                   + 18\ell^4)}{9(\rho+\ell^2)^5}\,,\nn\\
\bar S_3 &=& \fft{4\ell^4\, (5\rho^2 + 18\rho\, \ell^2 + 18\ell^4)}{
         9(\rho+\ell^2)^7}\,. \label{s3first}
\eea
 From (\ref{fsolution}) we find
\be
f = \fft{2\rho\, (21\rho^4 + 147 \rho^3\, \ell^2 + 
391 \rho^2\, \ell^4 + 471 \rho\, \ell^6 + 216 \ell^8)}{9
(2\rho + 3\ell^2)(\rho+\ell^2)^7}\,,\label{fexpli}
\ee
and so in the linearised level the perturbed solution is given by
(\ref{cfexp}), (\ref{barsolres}) and (\ref{fexpli}), where now $\varepsilon
= {\a'}^3/\ell^6$.   

    The function $f$ is non-singular in the entire coordinate range
$0\le \rho \le \infty$.  At large $\rho$ we have
\be
f = \fft{7}{3 \rho^3} -\fft{7\ell^2}{2\rho^2} +\cdots\,,
\ee
while at small $\rho$ we have
\be
f = \fft{16 \rho}{\ell^8} - \fft{790 \rho^2}{9\ell^{10}} + \cdots\,.
\ee
It is clear from this that the regularity of the metric on the $S^2$
bolt at $\rho=0$ is unaffected by the perturbation.  Of course since we
are working only to first-order in perturbations, it is necessary that
the parameter of the perturbation expansion be small compared to
unity.  The relevant dimensionless small parameter is $\a'/\ell^2$,
since $\ell$ sets the scale size of the bolt where the curvature of
the original metric reaches its maximum, at $\rho=0$.  In fact one can
see from (\ref{fexpli}) that $|f|$ reaches its maximum at about $\rho\sim
0.23 \ell^2$, with $|f|_{\rm max}$ being about $1.2\, \ell^{-6}$.  Thus
if $\a'/\ell^2$ is sufficiently small that the first-order perturbation
approximation is a good one, then the perturbed solution will be
non-singular everywhere.

    If one looks at the cone over $T^{1,1}$, corresponding to setting
the scale-size $\ell$ of the resolved conifold to zero, then 
ostensibly (\ref{s3first}) implies that $\bar S_3$ vanishes, suggesting
that the cone metric itself receives no modification from the corrections
at order $\alp$.  However, this is somewhat misleading since, as can
be seen from (\ref{s3first}), $\bar S_3$ reaches the value $8/\ell^6$
on the bolt in the resolved conifold, and thus it diverges in the limit
$\ell\longrightarrow 0$. Thus the assumption that $\a'/\ell^2$ is 
everywhere small is violated if one takes the $\ell\longrightarrow
0$ cone limit.  If $\ell$ is set equal to zero, $\bar S_3$ is still 
divergent at the apex of the cone, now with a delta-function behaviour.
Thus again it is strictly-speaking invalid to restrict attention to
only the $\alp$ corrections in this case.

   It is worth remarking that the {\it local} vanishing of $\bar S_3$
for the six-dimensional cone metric is an immediate consequence of the
fact that $\bar S_3$ is the bulk Euler integrand for six-dimensional
Ricci-flat metrics and hence $\int_0^\infty \bar S_3\, r^5\, dr$ must
be a finite number.  (The boundary contributions are discussed
in Appendix B.) A {\it generic} cubic invariant formed from the
Riemman tensor will have a $c/r^6$ power-law behaviour in the
six-dimensional cone metric, but the specific invariant in $\bar S_3$
must have $c=0$ since otherwise $\int_0^\infty \bar S_3\, r^5\, dr$
would be divergent.

          In terms of the comoving coordinate $t$, the functions $a$,
$b$ and $c$ have the following small-distance and large-distance behaviour:
\bea
\underline{t\rightarrow 0}\,: && \nn\\
&& a= \ft12 t\, \Big(1 - (\fft{1}{72} - \fft{4\alpha'^3}{3\ell^6})\,
\fft{t^2}{\ell^2} + \cdots\Big)\,,\nn\\
&& b= \ell\, \Big(1 + \fft{t^2}{8\ell^2} - (\fft{13}{1152} - 
\fft{\alpha'^3}{3\ell^6})\, \fft{t^4}{\ell^4} + \cdots\Big)\,,\nn\\
&& c= \ft12 t\, \Big(1 - (\fft{1}{18} - \fft{16\alpha'^3}{3\ell^6})\,
\fft{t^2}{\ell^2} + \cdots\Big)\,,\nn\\
\underline{t \rightarrow \infty}\,: &&\nn\\
&&a= \fft{t}{\sqrt6}\, \Big(1 - \fft{3\ell^2}{2t^2} + 
\fft{15\ell^4}{8t^4} - (\fft{207}{80} + \fft{504\alpha'^3}{5\ell^6})\,
\fft{\ell^6}{t^6} + \cdots\Big)\,,\nn\\
&&b= \fft{t}{\sqrt6}\, \Big(1 + \fft{3\ell^2}{2t^2} + 
\fft{15\ell^4}{8t^4} - (\fft{657}{80} + \fft{504\alpha'^3}{5\ell^6})\,
\fft{\ell^6}{t^6} + \cdots\Big)\,,\nn\\
&&c= \ft13t\, \Big( 1 - \fft{6\ell^4}{t^4} +
\ft{36}{5}\, (3 + \fft{\alpha'^3}{\ell^6})\, \fft{\ell^6}{\t^6}+
\cdots\Big)\,.
\eea

\subsection{Corrections to the deformed conifold}

   The deformed conifold is a second resolution of the conifold
metric, which has the topology of an $\R^3$ bundle over $S^3$.  It can
be written in the cohomogeneity-one form
\be
ds_6^2 = dt^2 + \ft14 a^2 \, [(\sigma_1-\Sigma_1)^2 + 
            (\sigma_2+\Sigma_2)^2] + \ft14 b^2\, [
 (\sigma_2-\Sigma_2)^2 + 
            (\sigma_1+\Sigma_1)^2] + \ft14 c^2\, (\Sigma_3-\sigma_3)^2\,,
\label{defmet}
\ee
for which we choose the vielbein basis
\bea
&&e^0=dt\,,\quad e^1=\ft12 a\, (\sigma_1-\Sigma_1)\,,\quad 
e^2= -\ft12 a\, (\sigma_2+\Sigma_2)\,,\nn\\
&&e^3=\ft12 b\, (\sigma_2-\Sigma_2)\,,\quad
e^4=\ft12 b\, (\sigma_1+\Sigma_1)\,,\quad
e^5=\ft12 c\, (\Sigma_3+\sigma_3)\,.
\eea

     The torsion-free spin connection is then summarised in the
vielbein components of the Lorentz-covariant exterior derivative
$\nabla = d + \ft14 \omega^{ab}\, \Gamma_{ab}$,
which we find to be
\bea
&&\nabla_0 = d_0\,,\nn\\
&&\nabla_1 = d_1  - \fft{\dot a}{2a}\, \Gamma_{01} 
   + \ft12 A\, \Gamma_{35} \,,\qquad
\nabla_2 = d_2  - \fft{\dot a}{2a}\, \Gamma_{02} 
   + \ft12 A\, \Gamma_{45} \,,\nn\\
&&\nabla_3 = d_3  - \fft{\dot b}{2b}\, \Gamma_{03} 
   - \ft12 B\, \Gamma_{15} \,,\qquad
\nabla_4 = d_4  - \fft{\dot b}{2b}\, \Gamma_{04} 
   - \ft12 B\, \Gamma_{25} \,,\nn\\
&&\nabla_5 = d_5  - \fft{\dot c}{2c}\, \Gamma_{05} 
   + \ft12 C\, (\Gamma_{13}+\Gamma_{24}) \,,
\label{defspincon}
\eea
where 
\be
A\equiv  \fft{a^2 -b^2-c^2}{2a\, b\, c}\,,\quad
B\equiv \fft{b^2 -a^2-c^2}{2a\, b\, c}\,,\quad
C\equiv \fft{c^2 -a^2-b^2}{2a\, b\, c}\,.\label{ABCdef}
\ee
There are additional terms $\omega_{12}^{\rm extra}= \omega_{34}^{\rm
extra} =-\ft12 (\Sigma_3+\sigma_3)$ that lie outside the $(S^3\times
S^3)/U(1)$ coset, and project to zero, as discussed in \cite{ricciflat}.
The Ricci curvature is found to be \cite{ricciflat}
\bea
R_{00} &=& -\fft{2\ddot a}{a} - \fft{2\ddot b}{b} -\fft{\ddot c}{c}
\,,\nn\\
R_{11} &=& R_{22} = -\fft{\ddot a}{a} - \fft{\dot a^2}{a^2} -
\fft{2\dot a\,\dot b}{a\,b} - \fft{\dot a\,\dot c}{a\,c} +
\fft{a^4 - b^4 - c^4 + 4 b^2\,c^2}{2a^2\,b^2\,c^2}\,,\nn\\
R_{33} &=& R_{44} = -\fft{\ddot b}{b} -\fft{\dot b^2}{b^2} -
\fft{2\dot a\,\dot b}{a\,b}-\fft{\dot b\, \dot c}{b\,c}
+ \fft{b^4 -a^4 -c^4 + 4a^2\, c^2}{2a^2\, b^2\, c^2}
\,,\nn\\
R_{55} &=& -\fft{\ddot c}{c} - \fft{2\dot a\,\dot b}{a\,b} -
\fft{\dot b\, \dot c}{b\,c} + \fft{c^4 - (a^2 -b^2)^2}{a^2\, b^2\, c^2}
\,.\label{defricci}
\eea

   It is evident from (\ref{defspincon}) that a spinor $\eta$ will be
covariantly constant if it has constant
components, satisfying the projection conditions
\be
\Gamma_{01}\eta= -\Gamma_{35}\eta\,,\qquad
\Gamma_{02}\eta= -\Gamma_{45}\eta\,,
\ee
provided that the first-order equations
\be
\dot a = - a\, A\,,\qquad \dot b = -b\, B\,,\qquad 
\dot c= - 2 c\, C\label{deffo}
\ee
hold.  These are the first-order equations whose solution yields the
Ricci-flat deformed conifold metric.  The K\"ahler form is given by
$J_{ab}=-\im\, \bar\eta\, \Gamma_{ab}\, \eta$, which gives
\be
J = - e^0\wedge e^5 + e^1\wedge e^3 + e^2\wedge e^4\,.
\ee

   If we now consider the corrected equation (\ref{ricmod}), then from 
(\ref{kseqn}) we see that only the component $\nabla_5$ receives a 
modification, namely the addition of $\ft{\im}{2} \dot S$, implying that
the corrected first-order equations become
\be
\dot a = - a\, A\,,\qquad \dot b = -b\, B\,,\qquad 
\dot c= - 2 c\, C - c\, \dot S \,,\label{deffo2}
\ee
where $A$, $B$ and $C$ are again given by (\ref{ABCdef}).  If these
equations are satisfied, then the metric (\ref{defmet}) will satisfy
the modified Einstein equations (\ref{ricmod}), which are
\be
R_{00}=R_{55}= \ddot S + \fft{\dot c}{c}\, \dot S\,,\qquad
R_{11}=R_{22}= \fft{2 \dot a}{a}\, \dot S\,,\qquad
R_{33}=R_{44}= \fft{2 \dot b}{b}\, \dot S\,,
\ee
where the Ricci tensor is given by (\ref{defricci}).

   Defining $u=a\, b$ and $v=a/b$, and introducing a new radial 
variable $r$ such that $dt=c\, dr$, the first-order equations (\ref{deffo2})
become 
\be
v'+ v^2 -1=0\,,\qquad u'=c^2\,,\qquad \fft{c'}{c} + \fft{c^2}{u} 
   - v -\fft{1}{v} + S'=0\,,
\ee
where a prime denotes a derivative with respect to $r$.  From these we can
solve to obtain 
\be
v=\coth r\,,\qquad u^3=\int_0^r e^{-2S}\,\sinh^2 2x\, dx\,,
\ee
and hence 
\be
a^2= u\, e^{S}\, \coth r\,,\quad
b^2= u\, e^{S}\, \tanh r\,,\quad
c = \fft{1}{\sqrt3 \, u}\, e^{-S}\, \sinh 2r\,.\label{defexact}
\ee
As in the case of the resolved conifold, if $S$ were an
externally-specified function then this would represent an exact
solution of the corrected first-order equations, and hence of the
corrected Einstein equations (\ref{ricmod}).  In our case $S$ is
itself an invariant constructed from the Riemann tensor, and so
(\ref{defexact}) is an integro-differential equation.

   Working to linear order in the perturbations, we can send
$S\longrightarrow \varepsilon\, S$, and write 
\be
a= \bar a\, (1+ \varepsilon\, f)\,,\qquad b= \bar b\, 
(1+\varepsilon\, f)\,,\qquad
c= \bar c\, (1+\varepsilon\, g)\,,\label{deffodef}
\ee
where the barred variables denote the metric functions in the
unperturbed Ricci-flat deformed conifold, and we work to linear order
in $\varepsilon$.  In fact the Ricci-flat deformed conifold solution
is given by
\be
\bar a= \ell \, R^{\ft16}\, (\coth r )^{\ft12}\,,\quad
\bar b = \ell\, R^{\ft16}\, (\tanh r)^{\ft12}\,,\quad
\bar c= \ft1{\sqrt3}\, \ell\, 
   R^{-\ft13}\, \sinh 2r \,,\label{defsol}
\ee
where $r$ is related to $t$ by $dt = \bar c\, dr$, and 
\be
R = \ft18 (\sinh 4 r   - 4r)\,.
\ee

   Substituting (\ref{deffodef}) into (\ref{defexact}), we now find
that the functions $f$ and $g$ are given by
\be
f= -\fft{\bar P}{\bar a^3\, \bar b^3} \,,\qquad 
g= \fft{2\bar P}{\bar a^3\, \bar b^3} - \bar S\,.\label{fgsol}
\ee
Here $\bar S$ denotes the curvature invariant appearing in 
(\ref{ssum}), evaluated in the unperturbed Ricci-flat metric.  The 
function $\bar P$ is then defined as
\be
\bar P(r) = \int_0^r\, dx\, \bar a(x)^2\, \bar b(x)^2\, 
    \bar c(x)^2\, \bar S(x)\,.
\ee

   If we take the special case of the $n=3$ term in (\ref{ssum}), then,
as we noted earlier, we can express $\sqrt{g}\, S_3$ as a total 
derivative.  In fact $\sqrt g$ is nothing but $a^2\, b^2\, c^2$ times
angular factors that are independent of the radial variable, and we
find that in this case we have 
\be
P = 96 a^2\, b^2\, c\, C\, [4 A^2 \, B^2 + (A+B)^2\, C^2]\,,
\ee
where $A$, $B$ and $C$ are defined in (\ref{ABCdef}).  Of course
one should replace $a$, $b$ and $c$ by their unperturbed expressions
(the barred variables) when substituting into (\ref{fgsol}).  

   Substituting the explicit expressions (\ref{defsol}) into
the first-order solution, we find that at short distance the perturbed
metric functions have the expansion
\bea
a &=& 2^{1/3}\, 3^{-1/6}\, \ell\, (1+ \ft{1152}{25}\varepsilon)\,
   \Big[ 1 + \fft{3(125-55296\varepsilon)\, r^2}{1250} +\cdots\Big]\,,\nn\\
b &=& 2^{1/3}\, 3^{-1/6}\, \ell\, (1+ \ft{1152}{25}\varepsilon)\,
   \Big[r - \fft{(125 +497664\varepsilon)\, r^3}{3750} +\cdots\Big]\,,\\
c &=& 2^{1/3}\, 3^{-1/6}\, \ell\, (1+ \ft{1152}{25}\varepsilon)\,
   \Big[ 1 + \fft{2(125-124416\varepsilon)\, r^2}{625} +\cdots\Big]\,,\nn
\eea
where $\varepsilon\equiv {\a'}^3/\ell^6$ here. At large distance, we
find the perturbed metric functions have the expansion
\bea
a &=& 2^{-2/3}\, \ell\, e^{\ft23 r}\, \Big[ 1 + e^{-2r} + 
                  \fft{9-24r + 5120\varepsilon}{18}\, e^{-4r} +\cdots
                     \Big]\,,\nn\\
b &=& 2^{-2/3}\, \ell\, e^{\ft23 r}\, \Big[ 1 + e^{-2r} + 
                  \fft{9-24r + 5120\varepsilon}{18}\, e^{-4r} +\cdots
                     \Big]\,,\\
c &=& 2^{1/3}3^{-1/2}\, \ell\, e^{\ft23 r}\, \Big[ 1 - 
                  \fft{9-24r + 5120\varepsilon}{18}\, e^{-4r} +\cdots
                     \Big]\,,\nn
\eea
We see that the effect of including the perturbation is to keep the
metric regular near the $S^3$ bolt at $r=0$, and provided the scale
size $\ell$ is large enough compared to $\sqrt{\a'}$, the metric will
be regular for all $r$.  Note, however, that the scale of the metric
is modified by a factor $(1 + 1152 {\a'}^3/(25\ell^6))$ at short
distance.  There was no analogous modification to the scale size of the
resolved conifold in Section \ref{resconsec}. 
   
\subsection{Corrections to the line bundle over $S^2\times S^2$}
\label{s2s2sec}

   The metric ansatz (\ref{rescon}) for the resolved conifold also
encompasses a different complete Ricci-flat metric, with a different
topology.  It corresponds to a situation where the principal orbits
degenerate to an $S^2\times S^2$ bolt rather than an $S^2$ bolt.  The
first-order equations remain the same as in (\ref{resfo}), with the 
same modified form (\ref{resfomod}) when the higher-order corrections
are turned on.  In fact the solutions now correspond simply to taking
both $\ell_1$ and $\ell_2$ to be non-zero in (\ref{abressol}), so
that neither $a$ nor $b$ vanishes as $\rho$ approaches zero. The
Ricci-flat solution is then given by \cite{ricciflat}
\be
a^2 = \rho +\ell_1^2\,,\qquad b^2=\rho+ \ell_2^2 \,,\qquad
\bar c^2 = \fft{\rho\, (2\rho^2 + 3(\ell_1^2 + \ell_2^2)\, \rho
       + 6 \ell_1^2\, \ell_2^2)}{3(\rho+\ell_1^2)(\rho+\ell_2^2)}\,.
\label{abcs2s2}
\ee
The topology of the principal orbits is changed also; one finds that
regularity of the metric at $\rho=0$ implies that the period of the
$U(1)$ fibre coordinate over $S^2\times S^2$ is now half of its value
in the $T^{1,1}$ orbits of the resolved conifold case, and so now the
principal orbits are $T^{1,1}/Z_2$.  The metric with $\ell_1=\ell_2$
was first given in \cite{berber,pagpop}.  For $\ell_1\ne\ell_2$, the
metric was given in \cite{pt1} in a different coordinate system.

   The analysis of the corrected solutions for this $\R^2$ bundle over
$S^2\times S^2$ is very similar to that for the resolved conifold in
Section \ref{resconsec}.  The only difference in the construction of

the linearly-perturbed solution is that now the constant $k$ in
(\ref{pres}) must be set to zero, in order to obtain a perturbed
solution that is regular at $\rho=0$.  The expressions for $\bar P$
and $\bar S_3$ are now rather complicated rational functions of
$\rho$, which we shall not present explicitly.  They are easily
constructed by substituting (\ref{abcs2s2}) into (\ref{pres}) and
(\ref{barS3}).  They are both finite everywhere, with asymptotic forms
\bea
\bar P &=& \fft{88}{9} - \fft{10(\ell_1^2-\ell_2^2)^2}{9\rho^2}  
+ \cdots\,,\nn\\
\bar S_3 &=& \fft{20(\ell_1^2-\ell_2^2)^2}{9\rho^5}
    - \fft{68(\ell_1^6 + \ell_2^6) -104 \ell_1^2\, \ell_2^2\, 
(\ell_1^2 + \ell_2^2)}{9\rho^6} + \cdots
\eea
at large distance, and
\bea
\bar P &=& \fft{36(\ell_1^2 + \ell_2^2)\, \rho}{\ell_1^2\, \ell_2^2}
  - \fft{6(15\ell_1^4 + 26 \ell_1^2\, \ell_2^2 + 15 \ell_2^4)\, \rho^2 
}{\ell_1^4\, \ell_2^4} + \cdots\,,\nn\\
\bar S_3 &=&  \fft{36(\ell_1^2 + \ell_2^2)}{\ell_1^4\, \ell_2^4}
- \fft{24(9\ell_1^4 + 16 \ell_1^2\, \ell_2^2 + 9\ell_2^4)\, \rho 
}{\ell_1^6\, \ell_2^6}+\cdots 
\eea
at short distance.  
Likewise the expression for $f$ given by (\ref{fsolution})
is quite involved, and so we shall just present its asymptotic forms
explicitly here.  It is finite everywhere, and at large distance we now find
\be
f = \fft{88}{3\rho^3} - \fft{44(\ell_1^2+\ell_2^2)}{\rho^4} + \cdots\,.
\ee
At small distance, we find
\be
f= \fft{12(9\ell_1^4 + 16 \ell_1^2\, \ell_2^2 + 9\ell_2^4)\, \rho}{
\ell_1^6\, \ell_2^6} - 
\fft{6(87\ell_1^6 + 211 \ell_1^4\, \ell_2^2 + 211 \ell_1^2\, \ell_2^4
+ 87 \ell_2^6)\, \rho^2}{\ell_1^8\, \ell_2^8} + \cdots\,.
\ee

    It is evident from these expressions that both $\ell_1$ and
$\ell_2$ must be non-vanishing for these perturbed solutions to be
regular.  In particular, this means that one cannot simply obtain the
modified solution for the resolved conifold by just setting $\ell_1=0$
in the modified solution for the $\R^2$ bundle over $S^2\times S^2$.
This is understandable, since we found that it was necessary to choose
$k=0$ rather then $k=-9$ in (\ref{pres}) in order to obtain a regular
modified solution for the $\R^2$ bundle over $S^2\times S^2$.

   A special case for the $\R^2$ bundle over $S^2\times S^2$ is when
$\ell_1=\ell_2$, implying that the $S^2\times S^2$ is itself an
Einstein metric.  The Ricci-flat solution is then encompassed in the
results of \cite{berber,pagpop}.  Since the functions in the perturbed
solution become much simpler in this case, we shall present them
explicitly here.  Setting $\ell_1=\ell_2=\ell$, we find
\bea
\bar P &=& \fft{88}{9} - \fft{8 \ell^6}{3(\rho+ \ell^2)^3} 
    - \fft{64 \ell^{18}}{9(\rho+\ell^2)^9}\,,\nn\\
\bar S_3 &=& \fft{8 \ell^6}{(\rho+\ell^2)^6} + \fft{64 \ell^{18}}{
   (\rho+\ell^2)^{12}}\,, \label{fl1eql2}\\
f&=& \fft{88}{3(\rho+\ell^2)^3} + \fft{40 \ell^6}{3(\rho+\ell^2)^6} 
     + \fft{64\ell^{12}}{3(\rho+\ell^2)^9} - 
          \fft{64 \ell^{18}}{(\rho+\ell^2)^{12}}\,.\nn
\eea

         In the comoving frame, the functions $a=b$ and $c$ have the following 
short-distance and large-distance behaviours:
\bea
\underline{t\rightarrow0}\,: &&\nn\\
&& a= \ell\, \Big( 1 + \fft{t^2}{4\ell^2} - (\fft{7}{96} -
\fft{34\alpha'^3}{\ell^6})\, \fft{t^4}{\ell^4} +
\cdots\Big)\,,\nn\\
&&c = t\,\Big(1 -(\ft13 - \fft{272\alpha'^3}{\ell^6})\,
\fft{t^2}{\ell^2} + \cdots\Big)\,,\nn\\
\underline{t\rightarrow\infty}\,: &&\nn\\
&&a= \fft{t}{\sqrt6}\, \Big( 1 + (\fft{108}5 -
\fft{6336\alpha'^3}{5\ell^6})\, \fft{\ell^6}{t^6} + \cdots\Big)\,,
\nn\\
&&c =  \ft13t\,\Big(1 - \ft{144}{5}(3 - \fft{176\alpha'^3}{\ell^6})\,
\fft{\ell^6}{t^6} + \cdots\Big)\,,
\eea

\subsection{Corrections to the line bundle over $\CP^2$}

    A final explicit Calabi-Yau example is provided by using the same
construction as for the $\R^2$ bundle over the Einstein metric on
$S^2\times S^2$, except that we now replace the $S^2\times S^2$ by the
Fubini-Study metric on $\CP^2$, with the same value for the
cosmological constant.  The Ricci-flat metric on the $\R^2$ bundle
over $\CP^2$ is again a special case of results in
\cite{berber,pagpop}.

   The metric can be written as
\be
ds_6^2 = dt^2 + 6 a^2\, d\Sigma_4^2 + c^2\, (dz +A)^2\,,\label{bbppcp2}
\ee
where $d\Sigma_4^2$ is the Fubini-Study metric on $\CP^2$, with its
canonical normalisation $R_{ij}= 6 g_{ij}$, and $dA$ is proportional
to its K\"ahler form.  We have included the factor of 6 in the
$d\Sigma_4^2$ term in (\ref{bbppcp2}) to scale the $\CP^2$ metric to
one with $R_{ij}= g_{ij}$, which is the same as we had for the
$S^2\times S^2$ base metric in Section \ref{s2s2sec}. The Fubini-Study
metric, and the potential $A$, can be written as \cite{gibpocp2}
\bea
d\Sigma_4^2 &=& d\xi^2 + \ft14 \sin^2\xi\, (\sigma_1^2 + \sigma_2^2)
        + \ft14 \sin^2\xi\, \cos^2\xi\, \sigma_3^2\,,\label{cp2metric}
\nn\\
A &=& -\ft32 \sin^2\xi\, \sigma_3\,,
\eea
where $\sigma_i$ are the left-invariant 1-forms of $SU(2)$.  It is 
straightforward to verify that the metric (\ref{bbppcp2}) is Ricci-flat
if the first-order equations
\be
\dot a = -\fft{c}{2a}\,,\qquad \dot c= -1 + \fft{c^2}{a^2}
\ee
hold.  As expected, these are the same as those for the $R^2$ bundle
over $S^2\times S^2$ with $a=b$. In terms of a new radial variable
$\rho$ such that $dt=- c^{-1}\, d\rho$, the solution is again given by
\be
a^2= \rho + \ell^2\,,\qquad c^2= \fft{2\rho\, (\rho^2 + 3\rho\, \ell^2 
                + 3\ell^4)}{3(\rho + \ell^2)^2}\,.
\ee
The regularity of the metric at small $\rho$ implies that the $U(1)$ fibre 
coordinate $z$ should have period $\Delta z=2\pi$, rather than the 
$\Delta z = 6\pi$ that would be required for $S^5$, and so the principal
orbits are $S^5/Z_3$.  
  
    Although the construction of the Ricci-flat K\"ahler metric is 
closely parallel to the case where the base is $S^2\times S^2$ rather 
than $\CP^2$, and the solution involves identical metric functions 
$a$ and $c$, we find that the effect of the ${\a'}^3$ corrections
is significantly different.  The essential point is that the Riemann
tensor for this $\R^2$ bundle over $\CP^2$ is different from that for
the $\R^2$ bundle over $S^2\times S^2$, since the Riemann tensors of the
four-dimensional bases are different, and hence the functional forms
of the Riemann-tensor invariants $S$ differ in the two cases.

   The modified first-order equations are of the identical form to
(\ref{primefo}), with $b=a$;
\be
a'=\fft{1}{a}\,,\qquad c' = \fft{1}{c} - \fft{c}{a^2} - c\, S'\,,
\ee
and so for the deformed solution we find
\be
a^2 =\rho +\ell^2\,,\qquad c^2 = \fft{2}{a^4}\, e^{-2S}\, \int_0^\rho
a(x)^4\, e^{2S(x)}\, dx\,.
\ee
Perturbatively, we send $S\longrightarrow \varepsilon\, S$, and 
write $c=\bar c\, (1 + \varepsilon\, f)$, finding
\be
f= \fft{2 \bar P}{a^4\, \bar c^2} - \bar S\,,\qquad \bar P = 
\int_0^\rho a(x)^4\, \bar S(x)\, dx\,,
\ee
precisely analogously to (\ref{fsolution}).  However, now we find that
$\bar P_3$ is given by
\be
\bar P_3 = -\fft{\bar c^2\, (4 a^4 - 6 a^2\, \bar c^2 + 3\bar c^4)}{
                                a^6}\,,
\ee
rather than (\ref{pres}).  In fact $\bar S_3$ itself now has a much
simpler form too, and is given simply by
\be
\bar S_3 = -\fft{1}{a^4}\, \fft{d\bar P_3}{d\rho} = 
\fft{8\bar c\, (2 a^2 - 3 \bar c^2)^3}{a^8}\,.
\ee
The function $f$ is now given by
\be
f = \fft{64}{3(\rho+\ell^2)^3} +  \fft{64\ell^6}{3(\rho+\ell^2)^6} +  
\fft{64\ell^{12}}{3(\rho+\ell^2)^9} -\fft{64\ell^{18}}{(\rho+\ell^2)^{12}}
\,.
\ee
It again has the property of vanishing at $\rho=0$ and $\rho=\infty$,
but it differs in detail from the result for $f$ in (\ref{fl1eql2})
for the case of the $\R^2$ bundle over $\R^2$.

      In terms of the comoving coordinate $t$, the function $a$ and $c$
have the following behaviour:
\bea
\underline{t\rightarrow0}\,: &&\nn\\
&&a=\ell\, \Big(1 + \fft{t^2}{4\ell^2} + (-\fft{7}{96} +
\fft{32\alpha'^3}{\ell^6})\,\fft{t^4}{\ell^4} + \cdots\Big)\,,
\nn\\
&&c= t\, \Big(1 + (-\ft13 + \fft{256\alpha'^3}{\ell^8})
\, \fft{t^2}{\ell^2} + \cdots\Big)\,,\nn\\
\underline{t\rightarrow\infty}\,:&&\nn\\
&&a= \fft{t}{\sqrt6}\, \Big(1 + (\fft{108}{5} -
\fft{4608\alpha'^3}{5\ell^6})\, \fft{\ell^6}{t^6} +
\cdots\Big)\,,\nn\\
&&c= \ft13 t\Big( 1 + (-\fft{432}{5} + \fft{18432\alpha'^3}{\ell^6})
\,\fft{\ell^6}{t^6} + \cdots\Big)\,.
\eea

\subsection{Corrections beyond ${\a'}^3$ order}

    Candidate correction terms at orders ${\a'}^4$ and above, consistent
with the highly-restrictive conditions of universality, were proposed
in \cite{frposost}.  They are given by (\ref{ricmod}) and (\ref{ssum}),
with
\be
S_n=R_{r_1\, k_2}{}^{r_2\, k_2}\, R_{r_2\, k_2}{}^{r_3\, k_3}\cdots
R_{r_n\, k_n}{}R^{r_1\, k_1} - 2^{n-2}\, 
R_{r_1}{}^{r_2}{}_{k_1}{}^{k_2}\,
R_{r_2}{}^{r_3}{}_{k_2}{}^{k_3}\cdots
R_{r_n}{}^{r_1}{}_{k_n}{}^{k_1}\,.\label{highod}
\ee
Clearly, up to and including order $\alpha'^5$, one can still use the
Ricci-flat background for calculating $S_n$, as we did for $S_3$.  
Interestingly, all the $S_n$ vanish for the
conifold itself, leading us to conjecture that the
conifold does not receive any higher-order corrections.  The vanishing of
$S_n$ for the conifold is non-trivial; it requires the precise relative 
values of the coefficients of the two terms in (\ref{highod}) that were 
conjectured in \cite{frposost}.

\bigskip
\noindent{\underline{Resolved conifold}}
\bigskip

        For the resolved conifold, we have
\bea
S_4&=&\fft{4\ell^4\,(14\rho^4 + 86\rho^3\,\ell^2 +
209\rho^2\,\ell^4 + 246\rho\,\ell^6 + 123\ell^8)}{
27(\rho + \ell^2)^{10}}\,,\\
S_5&=&\fft{40\ell^4\, (32\rho^6 + 277\rho^5\,\ell^2 +
1032\rho^4\,\ell^4 + 2140\rho^3\,\ell^6 +
2645\rho^2\, \ell^8 + 1890\rho\, \ell^{10} +
630\ell^{12})}{243(\rho + \ell^2)^{13}}\,,\nn
\eea
The corresponding correction to the function $c$ is given by
\be
c=\bar c \, (1 + \alpha'^3\, f + \alpha'^4\, f_4 + \alpha'^5\, f_5 +
\cdots)\,,
\ee
where $f$ was given in the previous subsection, and $f_4$ and $f_5$
are given by
\bea
f_4&=&\fft{2\rho}{189\ell^2\, (\rho + \ell^2)^{10}\,
(2\rho + 3\ell^2)}\,\Big(403\rho^7 + 4030\rho^6\, \ell^2 + 18135\rho^5\,
\ell^4 + 47576\rho^4\, \ell^6\nn\\
&&\qquad\qquad + 78554\rho^3\,\ell^8 + 81760\rho^2\,\ell^{10} +
49854\rho\,\ell^{12} + 13776\ell^{14}\Big)\,,\nn\\
f_5&=& \fft{2\rho}{18711\ell^4\, (\rho + \ell^2)^{13}\,
(2\rho + 3\ell^2)}\,\Big(120047 \rho^{10} + 1560611\rho^9\,\ell^2 +
9363666\rho^8\, \ell^4\nn\\
&&\qquad +34333442\rho^7\,\ell^6 +
85661125\rho^6\,\ell^8 _ 152736265\rho^5\,\ell^{10} +
198033000\rho^4\, \ell^{12}\nn\\
&&\qquad + 185217340\rho^3\,\ell^{14} +
120208550\rho^2\,\ell^{16} + 49191450\rho\,\ell^{18} +
9702000\ell^{20}\Big)\,,
\eea

         Thus we see that the higher-order corrections up to
$\alpha'^5$ all vanish at $\rho=0$ and $\rho=\infty$.  At small
$\rho$, the function $c$ takes the form
\be
c=\bar c\, \left(1 + \fft{16\alpha'^3\,(1750\alpha'^2 + 246\alpha'\,\ell +
81\ell^4)}{82\ell^{12}}\,\rho + {\cal O}(\rho^2)\right)\,,
\ee
whilst at large $\rho$ it takes the form 
\be
c=\bar c\, \Big(1 + \alpha'^3\, (\ft73 + \fft{403\alpha'}{189\ell^2} +
\fft{120047\alpha'^2}{18711\ell^4})\rho^{-3} + {\cal O}(\rho^{-4})
\Big)\,.
\ee
It is interesting to note that the coefficient of a given power of $\rho$
in these expansions receives corrections at each of the higher orders 
in $\a'$.  In the asymptotic region, it is
instructive to write the functions $a$, $b$ and $c$ in terms of a
comoving $t$ coordinate, in order to compare the asymptotic deviations of the
resolved conifold and the higher-order-corrected resolved conifold
from the cone metric itself.  These functions are given by
\bea
a\!\!\!\!&=&\!\!\!\!\fft{t}{\sqrt6}\, \Big(1 -\fft{3\ell^2}{2t^2} + 
\fft{15\ell^4}{8 t^{4}}
-(\ft{207}{80}\ell^6  + \ft{504}5\alpha'^3 + 
\ft{3224}{35}\alpha'^4\,\ell^{-2} +
\ft{15366016}{143451}\alpha'^5\,\ell^{-4})\, t^{-6} + \cdots\Big)\,,
\nn\\
b\!\!\!\!&=&\!\!\!\!\fft{t}{\sqrt6}\, \Big(1 +\fft{3\ell^2}{2t^{2}} + 
\ft{15\ell^4}{8 t^{4}}
-(\ft{657}{80} \ell^6 + \ft{504}5\alpha'^3 + 
\ft{3224}{35}\alpha'^4\,\ell^{-2} +
\ft{15366016}{143451}\alpha'^5\,\ell^{-4})\, t^{-6} + \cdots\Big)\,,
\nn\\
c\!\!\!\!&=\!\!\!\!&\ft13t\,\Big(1 -\fft{6\ell^4}{t^{4}} 
+ (\ft{108}{5}\ell^6 +
\ft{2016}{5}\alpha'^3 +\ft{12896}{35}\alpha'^4\,\ell^{-2} +
\ft{2841504}{3465}\alpha'^5\,\ell^{-4})t^{-6} + \cdots \Big)\,,
\eea
Thus we see that the higher-order corrections modify the asymptotic
behaviour in a rather mild fashion, and in particular, they are
highly normalisable at large distances.  It is interesting to note 
that purely on
dimensional grounds, one might have expected that $S_3$ could lead to
corrections of the form 
\be
{\a'}^3\,t^{-6}\,  \log t\,,
\ee
and in fact had the relative coefficient between the two terms in $S_3$
given in (\ref{s3exp}) been different, such a term would indeed arise.
Thus specific features of the actual higher-order corrections lead to
the systematic absence of structures in the series expansions.

\bigskip
\noindent{\underline{Deformed conifold}}
\bigskip

     For the deformed conifold, the explicit expressions for $S_4$ and
$S_5$ are rather complicated, and we shall not present them in detail.
The upshot is that the higher-order corrections have effects very
similar to the $\alpha'^3$ correction in modifying the small and large
distance behaviour.  To see this, we note that at small distances,
$S_3$, $S_4$ and $S_5$ are similar:
\bea
S_3 &=& \ft{3456}{125} (-5 + 24r^2 + \cdots)\,,\nn\\
S_4 &=& \ft{14976 6^{1/3}}{125} (-5 + 32 r^2 + \cdots)\,,\nn\\
S_5 &=& \ft{1760256 6^{2/3}}{625} (-1 + 8 r^2 + \cdots)\,,
\eea
whilst at large distances they all vanish.  Clearly the leading order
correction terms at large distance or at small distance will be
determined by the small distance behaviour for the $S_n$, which in
this case are all of the same form.  It follows that the corrections to
the deformed conifold have the same structure as those for the
$\alpha'^3$ correction, which we have already discussed.  

\bigskip
\noindent{\underline{$U(1)$ bundle over $S^2\times S^2$}}
\bigskip

      For the generic solution, the situation is rather similar,
but the structure of the solution is too complex to present here.  We
shall present only the case with $\ell_1=\ell_2\equiv\ell$, i.e, the
case considered in \cite{berber,pagpop}.  We have
\bea
S_4&=& \fft{64\ell^6}{9(\rho + \ell^2)^7} -
\fft{32\ell^{12}}{3(\rho + \ell^2)^{10}} +
\fft{2624\ell^{24}}{9(\rho + \ell^2)^{16}}\,,\nn\\
S_5&=&\fft{160\ell^6}{9(\rho + \ell^2)^8}
-\fft{320\ell^{12}}{27(\rho +\ell^2)^{11}} +
\fft{320\ell^{18}}{9(\rho + \ell^2)^{14}} +
\fft{89599\ell^{30}}{27(\rho + \ell^2)^{20}}\,,
\eea
The corresponding $f_4$ and $f_5$ are given by
\bea
f_4&=&\fft{6192}{91r\,\ell^6} + \fft{6192}{91r^4} +
\fft{45536\ell^6}{r^7} + \fft{3040\ell^{12}}{39r^{10}} +
\fft{2624\ell^{18}}{39 r^{13}} - \fft{2624\ell^{24}}{9r^{16}}\nn\\
&&-\fft{6292(r+\ell^2)}{91\ell^6\, (r^2 + r\,\ell^2 + \ell^4)}\,,\nn\\
f_5&=&\fft{112488}{187r^2\, \ell^2} + \fft{112488}{187r^5} +
\fft{964520\ell^6}{r^8} + \fft{3065600\ell^{12}}{5049r^{11}} +
\fft{84160\ell^{18}}{153r^{14}}\nn\\
&& +\fft{89600\ell^{24}}{153r^{17}} -\fft{89600\ell^{30}}{27 r^{20}}
-\fft{112488}{187\ell^6\, (r^2 + r\, \ell^2 + \ell^4)}\,.
\eea
where $r=\rho^2 + \ell^2$.  Again, the higher-order corrections vanish
for both $\rho=0$ and $\rho=\infty$.  However, the corrections have
more importance than in the previous conifold example.  In particular,
this is the case if we look at the large-distance behaviour.  Using the
comoving $t$ coordinate, we have
\bea
a&=&b=\fft{t}{\sqrt6}\,\Big( 1 + (\ft{108}5\ell^6 -
\ft{6336}{5}\alpha'^3 - \fft{1337472}{455}\alpha'^4\, \ell^{-2} -
\ft{24297408\alpha'^5}{935\ell^4})\, t^{-6} + \cdots\Big)\,,\nn\\
c&=&\ft13t\, \Big( 1 + (-\ft{432}{5}\ell^6 +
\ft{25344}{5}\alpha'^3 +
\fft{5349888}{455}\alpha'^4\, \ell^{-2} +
 \ft{97189632}{935}\alpha'^5\,\ell^{-4})\, t^{-6} + \cdots\Big)\,.
\eea
Thus we see that, in this case, the next-to-leading order terms in 
the conifold expansion are modified by higher-order corrections.

\bigskip
\noindent{\underline{$U(1)$ bundle over $\CP^2$}}
\bigskip

      In this case, the high-order correction sources $S_4$ and $S_5$
are rather simple; they are given by
\bea
S_4=\fft{164X^4}{9a^8}\,,\qquad
S_5=\fft{2800X^5}{27a^{10}}\,,
\eea
where $X=(2a^2 - 3 \bar c^2)/a^2$. It follows straightforwardly that
\bea
f_4 &=& \fft{2624}{117}\, \Big(\fft{3}{\ell^6\, r} +
\fft{3}{r^4} + \fft{3\ell^6}{r^7} + \fft{3\ell^{12}}{r^{10}} +
\fft{3\ell^{18}}{r^{13}} -\fft{13\ell^{24}}{r^{16}} -
\fft{3(r + \ell^2)}{\ell^6\, (r^2 + \ell^2\, r + \ell^4)}\Big)
\,,\\
f_5&=&\fft{89600}{459}\Big(\fft{3}{r^5} + \fft{3\ell^6}{r^8} +
\fft{3\ell^{12}}{r^{11}} + \fft{3\ell^{18}}{r^{14}} +
\fft{3\ell^{24}}{r^{17}} -\fft{17\ell^{30}}{r^{20}} +
\fft{3(r + \ell^2)}{\ell^4\, r^2\, (r^2 + r\, \ell^2 + \ell^4)}\Big)\,,\nn
\eea
where $r=\rho-\ell^2$.  In terms of the comoving coordinate $t$, the
functions $a$ and $f$ have the following behaviour
\bea
\underline{t\rightarrow0}\,: &&\nn\\
&& a=\ell\, \Big(1 + \fft{t^2}{4\ell^2} + (-\fft{7}{96} +
\fft{32\alpha'^3}{\ell^6} + \fft{5248\alpha'^4}{\ell^8} +
\fft{22400\alpha'^5}{\ell^{10}})\,\fft{t^4}{\ell^4} + \cdots\Big)\,,
\nn\\
&&c= t\, \Big(1 + (-\ft13 + \fft{256\alpha'^3}{\ell^8} +
\fft{41984\alpha'^4}{\ell^{8}} + \fft{1792000\alpha'^5}{81\ell^{10}}
)\, \fft{t^2}{\ell^2} + \cdots\Big)\,,\nn\\
\underline{t\rightarrow\infty}\,:&&\nn\\
&&a= \fft{t}{\sqrt6}\, \Big(1 + (\fft{108}{5} -
\fft{4608\alpha'^3}{5\ell^6} - \fft{188928\alpha'^4}{65\ell^8} -
\fft{430080\alpha'^5}{17\ell^{10}})\, \fft{\ell^6}{t^6} +
\cdots\Big)\,,\nn\\
&&c= \ft13 t\Big( 1 + (-\fft{432}{5} + \fft{18432\alpha'^3}{\ell^6} +
\fft{755712\alpha'^4}{65\ell^8} + \fft{1720320\alpha'^5}{17\ell^{10}})
\,\fft{\ell^6}{t^6} + \cdots\Big)\,.
\eea

\section{Explicit Non-compact Calabi-Yau Examples in $D=8$}\label{cy8sec}

   In this section, we investigate the effects of ${\a'}^3$ and higher
corrections on the various explicit examples of eight-dimensional 
non-compact Ricci-flat K\"ahler metrics of cohomogeneity one.  
These include the cases where the principal orbits are $U(1)$ 
bundles over $S^2\times S^2\times S^2$, $S^2\times \CP^2$ or $\CP^3$,
and the eight-dimensional Stenzel metric, for which the principal
orbits are $SO(5)/SO(3)$.

\subsection{$U(1)$ bundles over $S^2\times S^2\times S^2$}
\label{s2s2s2sec}

   We shall represent these metrics in terms of three
sets of left-invariant 1-forms for the group $SU(2)$, denoted by
$\sigma_i$, $\Sigma_i$ and $\nu_i$.  The eight-dimensional metric
is then given by
\be
ds_8^2 = dt^2 + a^2\, (\sigma_1^2+\sigma_2^2) +
b^2\, (\Sigma_1^2+\Sigma_2^2) +
c^2\, (\nu_1^2+\nu_2^2) + g^2\, (\sigma_3+\Sigma_3+\nu_3)^2\,.
\ee
We introduce the natural vielbein basis
\bea
&&e^0=dt\,,\quad e^1= a\, \sigma_1\,,\quad 
e^2= a\, \sigma_2\,,\quad  e^3= b\, \Sigma_1\,,
\quad e^4= b\, \Sigma_2\,,\nn\\
&& e^5= c\, \nu_1\,,\quad e^6= c\, \nu_2\,,\quad e^7= g\, 
(\sigma_3 + \Sigma_3 + \nu_3)\,.
\eea
Note that the two combinations $L_1\equiv \sigma_3-\nu_3$ and $L_2\equiv 
\Sigma_3 -\nu_3$ lie outside the coset.

   The torsion-free spin connection can be summarised in the expression
for the spinor-covariant exterior derivative $\nabla\equiv e^a\, \nabla_a
= d + \ft12 \omega_{ab}\, \Gamma^{ab}$,
\bea
&&\nabla_0 = d_0\,,\nn\\
&&\nabla_1 = d_1 - \fft{\dot a}{2a}\, \Gamma_{01} - \fft{g}{4 a^2}\, 
\Gamma_{27}\,,\qquad
\nabla_2 = d_2 - \fft{\dot a}{2a}\, \Gamma_{02} + \fft{g}{4 a^2}\, 
\Gamma_{17}\,,\nn\\
&&\nabla_3 = d_3 - \fft{\dot b}{2b}\, \Gamma_{03} - \fft{g}{4 b^2}\, 
\Gamma_{47}\,,\qquad
\nabla_4 = d_4 - \fft{\dot b}{2b}\, \Gamma_{04} + \fft{g}{4 b^2}\, 
\Gamma_{37}\,,\nn\\
&&\nabla_5 = d_5 - \fft{\dot c}{2c}\, \Gamma_{05} - \fft{g}{4 c^2}\, 
\Gamma_{67}\,,\qquad
\nabla_6 = d_6 - \fft{\dot c}{2c}\, \Gamma_{06} + \fft{g}{4 c^2}\, 
\Gamma_{57}\,,\nn\\
&&\nabla_7= d_7 - \fft{\dot g}{2g}\, \Gamma_{07} + 
\Big(\fft{g}{4a^2} -\fft1{6 g}\Big)\, \Gamma_{12} +
\Big(\fft{g}{4b^2} -\fft1{6 g}\Big)\, \Gamma_{34} +
\Big(\fft{g}{4c^2} -\fft1{6 g}\Big)\, \Gamma_{56}\,.
\eea
(There are also extra contributions $\omega_{12}^{\rm extra} = \ft13 
(L_2 - 2 L_1)$, $\omega_{34}^{\rm extra} = \ft13( L_1 - 2L_2)$, 
$\omega_{56}^{\rm extra} = \ft13 (L_1 + L_2)$, involving the two directions
outside the coset; these project out as discussed in \cite{ricciflat}.)

   It is easily seen that we can find two Killing spinors $\eta$ satisfying
$\nabla\, \eta=0$ which imply the first-order bosonic equations
\be
2 a\, \dot a = 2b\, \dot b = 2c\, \dot c = g\,,\qquad
\dot g = 1 - \ft12 g^2\, \Big(\fft{1}{a^2} + \fft1{b^2} + \fft1{c^2}
\Big)\,.
\ee
The spinors $\eta$ have constant components, and satisfy the projection
conditions
\be
\Gamma_{12}\, \eta = \Gamma_{34}\, \eta = \Gamma_{56}\, \eta = 
-\Gamma_{07}\, \eta\,.
\ee
The K\"ahler form can be written as 
$J_{ab} = -\im\, \bar\eta\, \Gamma_{ab}\, \eta$, and is given by
\be
J = e^1\wedge e^2 + e^3 \wedge e^4 + e^5 \wedge e^6 - e^0 \wedge e^7\,.
\ee

     The first-order equations that arise as integrability 
conditions for the modified Killing-spinor equation (\ref{kseqn}) are 
easily seen to be given by
\be
\dot a = \fft{g}{2a}\,,\qquad \dot b = \fft{g}{2b} \,,\qquad
\dot c = \fft{g}{2c}\,,\qquad
\dot g = 1 - \fft{g^2}{2a^2} - \fft{g^2}{2b^2} -\fft{g^2}{2 c^2} 
-g\, \dot S\,.
\label{d8resfomod}
\ee
As in the previous examples, one can easily verify that if these equations
are satisfied then the Einstein equations $R_{ab} = \nabla_a\nabla_b S
+ \nabla_{\hat a}\nabla_{\hat b} S$ are satisfied, where as usual
$\nabla_{\hat a} \equiv J_{a}{}^b\, \nabla_b$. These second-order
equations are
\be
R_{00}=R_{77} = \ddot S + \fft{\dot g}{g}\, \dot S\,,\quad
R_{11}=R_{22} = \fft{2 \dot a}{a}\, \dot S\,,\quad
R_{33}=R_{44} = \fft{2 \dot b}{b}\, \dot S\,,\quad
R_{55}=R_{66} = \fft{2 \dot c}{c}\, \dot S\,,
\ee
where the Ricci tensor is 
given by
\bea
R_{00} &=& -\fft{2\ddot a}{a} -\fft{2\ddot b}{b} - \fft{2\ddot c}{c} -
\fft{\ddot g}{g}\,,\nn\\
R_{11} &=& R_{22} = -\fft{\ddot a}{a} -\fft{\dot a^2}{a^2} -
\fft{2\dot a\, \dot b}{a\,b} - \fft{2\dot a\, \dot c}{a\, c} -
\fft{\dot a\, \dot g}{a\, g} + \fft{2a^2-g^2}{2a^4}\,,\nn\\
R_{33}&=&R_{44} =-\fft{\ddot b}{b} -\fft{\dot b^2}{b^2} -
\fft{2\dot a\, \dot b}{a\,b} - \fft{2\dot b\, \dot c}{b\, c} -
\fft{\dot b\, \dot g}{b\, g} + \fft{2b^2-g^2}{2b^4}\,,\nn\\
R_{55}&=&R_{66}=-\fft{\ddot c}{c} -\fft{\dot c^2}{c^2} -
\fft{2\dot a\, \dot c}{a\,c} - \fft{2\dot b\, \dot c}{b\, c} -
\fft{\dot c\, \dot g}{c\, g} + \fft{2c^2-g^2}{2c^4}\,,\nn\\
R_{77}&=&-\fft{\ddot g}{g} - \fft{2\dot g}{g}\, (\fft{\dot a}{a} +
\fft{\dot b}{b} + \fft{\dot c}{c}) + \ft12 g^2\, (\fft{1}{a^4} +
\fft{1}{b^4} + \fft{1}{c^4})\,.
\eea

   Introducing a new radial variable $\rho$ such that $d\rho = g\, dt$, 
it is easily seen that the solution to the modified first-order equations
(\ref{d8resfomod}) is given by
\bea
&&a^2 = \rho +\ell_1^2\,,\qquad
b^2 = \rho +\ell_2^2\,,\qquad
c^2 = \rho +\ell_3^2\,,\nn\\
&& g^2 = \fft{2}{a^2\, b^2\, c^2}\, e^{-2S}\, \int_0^\rho a^2(x)\, 
b^2(x)\, c^2(x)\, e^{2S(x)}\, dx\,.
\eea
As in our previous examples, this result is exact, and it is explicit  
(up to quadratures) if $S$ is a given externally-specified function.

   Our present interest is in the case where $S$ is some higher-order
correction term coming from string theory, as in the discussion of the
previous sections. We again therefore make a linearised approximation,
in which the quantity $S$ is expressed in terms of the background
Riemann tensors of the original Ricci-flat equations.  Sending
$S\longrightarrow \varepsilon\, S$, and writing $g = \bar g\, (1 +
\varepsilon\, f)$, where $\bar f$ is the expression for $f$ at zero'th
order in $\varepsilon$, we therefore find, up to linearised order,
that the metric functions are given by
\bea
&&a^2 = \rho +\ell_1^2\,,\qquad
b^2 = \rho +\ell_2^2\,,\qquad
c^2 = \rho +\ell_3^2\,,\nn\\
&& f_3 = \fft{2 \bar P}{a^2\, b^2\, c^2\, \bar f^2} - \bar S\,,
\eea
where
\be
 P_3 \equiv \int_0^\rho a^2(x)\, b^2(x)\, c^2(x)\, S(x)\, dx
\ee
and the quantities $\bar P$ and $\bar S$ are evaluated in the
zero'th-order Ricci-flat background.

     The general structure with $\ell_i$ not equal is rather
complicated to present.  We shall give explicit results only for the
case with $\ell_i=\ell$.  First let us consider the simplest case with
$\ell=0$, where the metric is just the cone over the $U(1)$ bundle
over $S^2\times S^2\times S^2$.  Unlike the six-dimensional conifold,
where $S_3$ vanishes locally (because, as we discussed, it is the 
six-dimensional Euler integrand), here it is non-vanishing and is given by
$S_3=3/\rho^3$.  It follows that the perturbation function $f_3$ is
given by $f_3=9\epsilon/\rho^3$.  This raises the possibility that the
string higher-order corrections might have the effect of resolving the
singularity of the cone metric itself.

        For $\ell\ne 0$, the perturbation function $f_3$ is given by
\be
f_3=\fft{105}{r^3} + \fft{153\ell^8}{2r^7} + \fft{90\ell^{16}}{r^{11}}
-\fft{495\ell^{24}}{2r^{15}} -
\fft{48}{\ell^4\,(r + \ell^2)} + \fft{48(r-\ell^2)}{\ell^4\,
(r^2 + \ell^4)}\,,
\ee
where $r=\rho-\ell^2$.
In the comoving $t$ coordinate, the asymptotic behaviour of the 
metric functions is
\bea
\underline{t\rightarrow 0}\,:&&\nn\\
&&a= \ell + \fft{t^2}{4\ell} - (\fft{3}{32\ell^3} -
\fft{159\epsilon}{\ell^9})\, t^4 + \cdots\,,\qquad
g=t - (\fft1{2\ell^2} - \fft{1272\epsilon}{\ell^8})\, t^3 +
\cdots\,,\nn\\
\underline{t\rightarrow \infty}\,:&&\nn\\
&&a= \fft{t}{2\sqrt2}\,\Big(1 -\fft{4608\epsilon}{5t^6} +
\fft{2048(\ell^8-192\ell^2\, \epsilon^2)}{7t^8} + \cdots\Big)\,,\nn\\
&&g= \ft14t\, \Big(1 + \fft{18432\epsilon}{5t^6} -
\fft{12288 (\ell^8-192\ell^2\, \epsilon)}{7t^8} +
\cdots\Big)\,.
\eea
Note that at large distances, the higher-order corrections modify
terms occurring before the next-to-leading order terms of 
the uncorrected expansion.  This is because, unlike in six dimensions, 
the integral of $S_3$ diverges at large distance in eight dimensions.

\subsection{$U(1)$ bundle over $S^2\times \CP^2$}

When $b=a$, we can replace the $S^2\times S^2$ with the metric for
$\CP^2$.  The metric ansatz now becomes
\be
ds^2= dt^2 + a^2 d\Sigma_4^2 + c^2\, d\Omega_2^2 + f^2\,
(dz -\ft32 \sin^2\xi\, \sigma_3 + A)^2\,,
\ee
where $dA=\Omega_\2$, and $d\Sigma_4^2$ is given by (\ref{cp2metric}).
At the zero'th order, the solutions for $a$, $c$ and $f$ are identical
to that of the previous case.  However, since the Riemann tensor for
$S^2\times S^2$ is different from that for $\CP^2$, it follows that
the $\alpha'^3$ correction term $S_3$ is different in this case from
the $S^2\times S^2\times S^2$ case.  For simplicity, we shall only
present the result when the constants are chosen so that
$a=c=\sqrt{\rho + \ell^2}$.  For the cone metric (\ie $\ell=0$) we now
find $S_3=1/r^3$ instead of $3/r^3$ for the $S^2\times S^2\times S^2$
case.  It follows that there are differences in the higher-order
corrections, but they are qualitatively the same.

   For $\ell\ne 0$, we find that
\be
f_3=\fft{99}{r^3} + \fft{165\ell^8}{2r^7} + 
\fft{90\ell^{16}}{r^{11}} - \fft{495\ell^{24}}{2r^{15}} -
\fft{48}{\ell^4\, (r + \ell^2)} + \fft{48(r-\ell^2)}{\ell\,
(r^2 + \ell^4)}\,.
\ee
where again $r=\rho-\ell^2$.  Thus structurally, the correction terms
are the same as those for the $U(1)$ bundle over $S^2\times S^2\times
S^2$, but the detailed coefficients are rather different, In the
comoving frame, at small distance $t$, $a$ and $f$ are given by
\bea
\underline{t\rightarrow 0}\,:&&\nn\\
&&a=\ell + \fft{t^2}{4\ell} - (\fft{3}{32\ell^3} -
\fft{157\epsilon}{\ell^9})\, t^4 + \cdots\,,\qquad
g=t - (\fft1{2\ell^2} - \fft{1256\epsilon}{\ell^8})\, t^3 +
\cdots\,,\nn\\
\underline{t\rightarrow \infty}\,:&&\nn\\
&&a=\fft{t}{2\sqrt2}\,\Big(1 -\fft{1536\epsilon}{5t^6} +
\fft{2048(\ell^8-192\ell^2\, \epsilon^2)}{7t^8} + \cdots\Big)\,,\nn\\
&&g= \ft14t\, \Big(1 + \fft{6144\epsilon}{5t^6} -
\fft{12288 (\ell^8-192\ell^2\, \epsilon)}{7t^8} +
\cdots\Big)\,,
\eea

\subsection{$U(1)$ bundle over $\CP^3$}

    When $a=b=c$, we can replace $S^2\times S^2\times S^2$ with 
$\CP^3$.  There are two convenient ways to write the $\CP^3$ metric.
One way is to use the recursive expression for the Fubini-Study metric
$d\Sigma_{2n}^2$ on $\CP^n$ in terms of the Fubini-Study metric 
$d\Sigma_{2n-2}^2$ on $\CP^{n-1}$, which was derived in \cite{hoxmarpop}:
\be
d\Sigma_{2n}^2 = d\a^2 + \sin^2\a\, d\Sigma_{2n-2}^2 + \sin^2\a\, \cos^2\a\,
(d\tau + B)^2\,,\label{cpnrecursion}
\ee
where $dB=2J_{n-1}$, and $J_{n-1}$ is the K\"ahler form of
$\CP^{n-1}$.  For each $n$, $d\Sigma_n^2$ denotes the
canonically-normalised Fubini-Study metric, with $R_{ij}= 2(n+1)\,
g_{ij}$.  Thus before taking the $O({\a'})$ corrections into account,
first-order equations for the metric
\be
ds_8^2 = dt^2 + 8 a^2\, d\Sigma_6^2 + g^2\, (dz + A)^2\label{bbppcp3}
\ee
will be the same as those for the $S^2\times S^2\times S^2$ base given
in Section \ref{s2s2s2sec}, with $a=b=c$.  The K\"ahler form for the
$\CP^n$ metric (\ref{cpnrecursion}) is given by $J_n = \ft12 dA$,
where $A=\sin^2\a\, (d\tau + B)$ \cite{hoxmarpop}.  Using
(\ref{cpnrecursion}) the metric on $\CP^3$ can be written in terms of
the $\CP^2$ metric (\ref{cp2metric}), with $A=-\ft14 \sin^2\xi\,
\sigma_3$.

    An alternative construction for the $\CP^n$ metrics can be given by
introducing left-invariant 1-forms $L_A{}^B$ for the group $SU(n+1)$,
where $0\le A\le n$, $L_A{}^A=0$, and $dL_A{}^B = \im\, L_A{}^C\wedge
L_C{}^B$.  Writing $A=(0,i)$, where $1\le i\le n$, we can obtain a vielbein
for the coset $\CP^n=SU(n+1)/U(n)$ by taking just the subset $L_0{}^i$ and 
$L_i{}^0$ of the left-invariant 1-forms, \ie by modding out by the 
$SU(n)$ 1-forms $L_i{}^j$ and $U(1)$ 1-form $L_0{}^0$.  In a real
basis, we can define
\be
e^i = \ft12(L_0{}^i + L_i{}^0),,\qquad e^{\td i} = \ft{1}{2\im}\, 
(L_0{}^i - L_i{}^0)\,.
\ee
The spin connection and curvature 2-forms for $\CP^n$ are therefore given by
\bea
&&\omega_{ij}= \omega_{\td i\td j} = \ft1{2\im}\, (L_i{}^j - L_j{}^i)\,,\qquad
\omega_{i\td j} = L_0{}^0\, \delta_{ij} - \ft12 (L_i{}^j+ L_j{}^i)\,,\nn\\
&&\Theta_{ij}=\Theta_{\td i\td j} = e^i\wedge e^j + e^{\td i}\wedge 
e^{\td j}\,,\qquad \Theta_{i\td j} = e^i\wedge e^{\td j} +
   e^j\wedge e^{\td i} + 2 e^k\wedge e^{\td k} \, \delta_{ij}\,,
\eea
and the K\"ahler form is
\be
J= e^i\wedge e^{\td i}\,.
\ee
Thus we see that $R_{ij}=2(n+1)\, \delta_{ij}$, and hence the metric
$d\Sigma_{2n}^2 = e^i\, e^i + e^{\td i}\, e^{\td i}$ is the 
canonically-normalised Fubini-Study metric on $\CP^n$.

   Using either of the above constructions, it is a straightforward matter
to calculate the curvature for the metric (\ref{bbppcp3}), 
and hence to show that the cubic Riemann tensor invariant is given 
in this case by 
\be
S_3=\fft{495(a^2-2g^2)^3}{2a^{12}}\,.
\ee
Clearly, the cone of the $U(1)$ bundle over $\CP^3$, corresponding
to $a=\sqrt2\, g$, is locally Euclidean since the principal orbits are 
locally the round $S^7$, and hence locally the curvature and all higher-order
corrections vanish.  If $\ell \ne 0$, we have
\be
f_3 = \fft{90}{r^3} + \fft{90\ell^8}{r^7} + \fft{90\ell^{16}}{r^{11}}
-\fft{495\ell^{24}}{2r^{15}} -\fft{45}{\ell^4\, (r + \ell^2)} -
\fft{45(r-\ell^2)}{\ell^4\, (r^2 + \ell^4)}\,,
\ee
where $r=\rho-\ell^2$.   
Since now $S_3$ is normalisable, the correction is very different from
the previous ones.  In the comoving frame, $a$
and $g$ have the asymptotic forms
\bea
\underline{t\rightarrow 0}\,:&&\nn\\
&&a=\ell + \fft{t^2}{4\ell} - (\fft{3}{32\ell^3} -
\fft{2475\epsilon}{16\ell^9})\, t^4 + \cdots\,,\qquad
g=t - (\fft1{2\ell^2} - \fft{2475\epsilon}{2\ell^8})\, t^3 +
\cdots\,,\nn\\
\underline{t\rightarrow \infty}\,:&&\nn\\
&&a=\fft{t}{2\sqrt2}\,\Big(1 +
\fft{2048(\ell^8-180\ell^2\, \epsilon^2)}{7t^8} + \cdots\Big)\,,\nn\\
&&g= \ft14t\, \Big(1 -
\fft{12288 (\ell^8 +30\ell^2\, \epsilon)}{7t^8} +
\cdots\Big)\,.
\eea

\subsection{Stenzel metrics; $SO(5)/SO(3)$ orbits}

  We shall closely follow the notation of \cite{ricciflat} for writing
the cohomogeneity one metrics with $SO(n+2)/SO(n)$ principal orbits:
\be
ds^2 = dt^2 + a^2\, \sigma_i^2 + b^2\, \td\sigma_i^2 + c^2\, \nu^2\,,
\label{stenzelmet}
\ee
where $3\le i \le n+2$, 
\be
\sigma_i\equiv L_{1i}\,,\qquad \td\sigma_i \equiv L_{2i}\,,\qquad
\nu\equiv L_{12}\,,
\ee
and the $L_{AB}$ with $1\le A\le n$ are the 
left-invariant 1-forms of the group $SO(n+2)$, satisfying $dL_{AB} = 
L_{AC}\wedge L_{CB}$, with $L_{AB}=-L_{BA}$.  We choose the natural 
orthonormal basis
\be
e^0 = dt\,,\qquad e^i = a\, \sigma_i\,,\qquad
 e^{\td i} = b\, \td\sigma_i\,,\qquad
e^{\td 0} = c\, \nu\,.\label{stenzelcon}
\ee

   We take $\bar e^0\equiv \nu$ and $\bar
e^i\equiv \sigma_i$ as a vielbein basis $\bar e^a$ for the sphere $S^{n+1} =
SO(n+2)/SO(n+1)$.  A simple calculation shows that the torsion-free
spin connection is given by $\bar \omega_{0i}= -\td\sigma_i$,
$\omega_{ij} = -L_{ij}$, and hence that the curvature 2-forms for the
metric $d\bar s^2\equiv \bar e^a\, \bar e^a= \sigma_i^2 + \nu^2$ are given by
$\bar \Theta_{ab} = e^a\wedge e^b$.  This proves that $d\bar s_{n}^2$
is the metric on the {\it unit} $(n+1)$-sphere.  Thus the $SO(n+2)/SO(n)$ 
principal orbits in (\ref{stenzelmet}) can be viewed as $S^n$ fibres over
a (squashed) $S^{n+1}$ base, with $\td\sigma_i$ being 1-forms on the
$S^n$ fibres.\footnote{The roles of the $\sigma_i$ and the $\td\sigma_i$ are
symmetrical in this description, and they could be interchanged.} 

  Calculating the torsion-free spin connection for (\ref{stenzelcon}),
one finds that the spinor covariant exterior derivative is given by
\bea
\nabla_0 &=& d_0\,,\nn\\
\nabla_i &=& d_i - \fft{\dot a}{2a}\, \Gamma_{0i} - \ft12 A\, 
\Gamma_{\td 0 \td i}\nn\\
\nabla_{\td i} &=& d_{\td i} - \fft{\dot b}{2b}\, \Gamma_{0\td i} 
 \ft12 B\, 
\Gamma_{\td 0 i}\nn\\
\nabla_{\td 0} &=& d_{\td 0} - \fft{\dot c}{2c}\, \Gamma_{0\td 0} + 
\ft12 C\, \Gamma_{i\td i}\,,\label{spincovs8}
\eea
where 
\be
A\equiv  \fft{a^2 -b^2-c^2}{2a\, b\, c}\,,\quad
B\equiv \fft{b^2 -a^2-c^2}{2a\, b\, c}\,,\quad
C\equiv \fft{c^2 -a^2-b^2}{2a\, b\, c}\,.\label{ABCdef8}
\ee
(There are also additional terms $\omega_{ij}^{\rm extra}=
\omega_{\td i \td j}^{\rm extra} = - L_{ij}$ that lie outside
the coset, and that project to zero \cite{ricciflat}.)  

   It is evident from these first-order equations that there is
a solution whose short-distance behaviour (near $t=0$) takes the form
\be
ds^2 = dt^2 + t^2\, \td\sigma_i^2 + a_0^2\, (\sigma_i^2 + \nu^2)\,.
\label{stenshort}
\ee
This is precisely the short-distance behaviour of the Stenzel metrics
\cite{ricciflat}, which are complete and non-singular.  It is clear from 
(\ref{stenshort}) that the metric $\td\sigma_i^2$ on the $S^n$ fibres
must describe a sphere of {\it unit} radius, in view of the regularity
at $t=0$.  Thus we can conclude that the principal $SO(n+2)/SO(n)$ orbits 
in the Stenzel metrics have a volume given by
\be
\hbox{Vol}(SO(n+2)/SO(n)) = \int \nu\wedge \prod_i \sigma_i\wedge
\prod_i \td\sigma_i = \hbox{Vol}(S^{n+1})\, \hbox{Vol}(S^n) \,.
\ee
Since the volume of the unit $n$-sphere is Vol$(S^n)=2 \pi^{(n+1)/2}/
\Gamma\big((n+1)/2\big)$, it follows that the volumes of the principal orbits 
in the Stenzel metrics are given by
\be
\hbox{Vol}(SO(n+2)/SO(n)) = \fft{2^{n+2}\, \pi^{n+1}}{n!}\,.
\ee
We shall make use of this result later, when calculating the 
contributions of the volume and boundary terms in the expression for the
Euler number.

  Specialising to $D=8$ (\ie $n=3$), the Ricci tensor is given by
\cite{ricciflat}
\bea
R_{00} &=& -\fft{3\ddot a}{a} - \fft{3\ddot b}{b} - \fft{\ddot c}{c}
\,,\nn\\
R_{11}&=&R_{22} =R_{33}=-\fft{\ddot a}{a} - \fft{2\dot a^2}{a^2} -
\fft{3\dot a\, \dot b}{a\,b} - \fft{\dot a\, \dot c}{a\, c} +
\fft{a^4 - b^4 - c^4 + 6b^2\, c^2}{2a^2\, b^2\, c^2}\,,\nn\\
R_{44}&=&R_{55}= R_{66}=-\fft{\ddot b}{b} - 
\fft{2\dot b^2}{b^2} -\fft{3\dot a\, \dot b}{a\,b} -
\fft{\dot b\,\dot c}{b\, c} +
\fft{b^4 - a^4 - c^4 + 6 a^2\, c^2}{2a^2\, b^2\, c^2}\,,\nn\\
R_{77}&=& -\fft{\ddot c}{c} - \fft{3 \dot a\, \dot c}{a\, c} -
\fft{3 \dot b\, \dot c}{b\, c} + 
\fft{3(c^2-(a^2 -b^2)^2)}{2a^2\, b^2\, c^2}\,.\label{sten8ricci}
\eea

   As in the previous examples, we can read off from the covariant exterior
derivative (\ref{spincovs8}) the first-order integrability conditions for
the existence of covariantly-constant spinors $\nabla\eta=0$, giving
\be
\dot a = - a\, A\,,\qquad \dot b = -b\, B\,,\qquad 
\dot c= - 3 c\, C \,,\label{d8deffo20}
\ee
where the spinors have constant components and satisfy the projection 
conditions
\be
\Gamma_{0i}\, \eta + \Gamma_{\td 0 \td i}\, \eta=0\,.
\ee
The K\"ahler form can be written as $J_{ab} = -\im\, \bar\eta \Gamma_{ab}\, 
\eta$, giving
\be
J= - e^0\wedge e^{\td 0} + e^i\wedge e^{\td i}\,.
\ee

  The integrability conditions for the 
modified Killing spinor equation (\ref{kseqn}) are then easily seen to be
\be
\dot a = - a\, A\,,\qquad \dot b = -b\, B\,,\qquad 
\dot c= - 3 c\, C - c\, \dot S \,.\label{d8deffo2}
\ee
As in the previous cases, one can verify that if these equations are
satisfied, then the metric satisfies the modified Einstein equations
$R_{ab} = \nabla_a\nabla_b S + \nabla_{\hat a}\nabla_{\hat b} S$.  
Explicitly, these equations are
\be
R_{00}= R_{\td 0 \td 0} = \ddot S + \fft{\dot c}{c}\, \dot S\,,\qquad
R_{ij} = \fft{2\dot a}{a}\, \dot S\, \delta_{ij}\,,\qquad
R_{\td i\td j} = \fft{2\dot b}{b}\, \dot S\, \delta_{\td i\td j}\,,
\ee
where the Ricci tensor is given by (\ref{sten8ricci}).

   The first-order equations (\ref{d8deffo2}) can be solved by
defining $u\equiv a\, b$, $v\equiv a/b$ and introducing a new radial
variable $r$ such that $dt =c\, dr$.  The first-order equations become
\be
v' + v^2 -1=0\,,\qquad u'=c^2\,,\qquad \fft{c'}{c} + \fft{3 u'}{2u} 
 -\ft32 (v + v^{-1}) + S'=0\,,
\ee
leading to the solution
\be
v=\coth r\,,\qquad u^4 = \int_0^r e^{-2S(x)}\, (\sinh 2x)^3\, dx\,,
\qquad c^2 = \ft14 e^{-2S}\, u^{-3}\, (\sinh 2r)^3 \,.
\ee

   In our perturbative discussion, we can solve explicitly for the
linearised deformations by sending $S\longrightarrow \varepsilon \, S$,
and writing 
\be
a= \bar a\, (1 + \varepsilon\, f)\,,\qquad
b= \bar b\, (1 + \varepsilon\, f)\,,\qquad 
c = \bar c\, (1 + \varepsilon\, g)\,,
\ee
where the barred quantities denote the zero'th-order Ricci-flat 
expressions.  These are given by \cite{ricciflat}
\be
\bar a^2 = R^{1/4}\, \coth r\,,\qquad \bar b^2 = R^{1/4}\, \tanh r
\,,\qquad \bar c^2 = \ft14 R^{-3/4}\, (\sinh 2r)^3\,,
\ee
where 
\be
R \equiv \ft32 ( 2 + \cosh 2r)\, \sinh^4 r\,.
\ee
Solving for $f$ and $g$ at linearised order in $\varepsilon$, we then
find
\be
f = - \fft{\bar P}{\bar a^4\, \bar b^4}\,,\qquad
g = \fft{3\bar P}{\bar a^4\, \bar b^4} - \bar S\,,\label{fgsols}
\ee
where
\be
P \equiv  \int_0^r a^3(x)\, b^3(x)\, c^2(x)\, S(x)\, dx\,,
\ee
and the quantities $\bar P$ and $\bar S$ are evaluated using the Riemann
tensor in the undeformed Ricci-flat background.

        For the $\alpha'^3$ corrections, we find that $f_3$ and $g_3$
are given by
\bea
f_3 &=& \fft{80\, 3^{\ft34}\, e^{\ft{37}2 r}}{(1 + e^{2r})^{11}\,
(1 + 4e^{2r} + e^{4r})^{\ft{15}4}}\, \Big(
272488 + 434591 \cosh{2 r} + 225766 \cosh{4 r}\nn\\
&&+ 78287 \cosh{6 r} +18120 \cosh{8 r} + 2697 
\cosh{10 r} + 234 \cosh{12 r} + 9 \cosh{14 r}\Big)\,,\nn\\
g_3 &=& \fft{160 e^{\ft{45}2 r}}{3^{1/3} (1 + e^{2r})^{15}\, 
(1 + 4e^{2r} + e^{4r})^{\ft{15}4}}\, \Big(
9352650 + 13111232 \cosh{2r} + 3993140 \cosh{4 r}\nn\\
&&- 415614 \cosh{6 r} - 835680 \cosh{8 r} - 343762 \cosh{10 r}
- 77940 \cosh{12 r}\nn\\
&& - 10581 \cosh{14 r} - 810 \cosh{16 r} - 
27 \cosh{18 r}\Big)\,,
\eea
In the comoving coordinate $t$, the metric functions $a$, $b$ and $c$
behave in the following way at small distances
\bea
a&=& \Big(2^{\ft18} + \fft{280\, 2^{3/8}\, \epsilon}{3}\Big) \Big(
1 + \ft16 (2^{3/4} - 2240\epsilon)\, t^2 + \cdots\Big)\,,\nn\\
b&=& t\, \Big (1 - (\fft1{6\, 2^{1/4}} - \fft{280\epsilon}9)\, t^2 +
\cdots\Big)\,,\nn\\
c&=& \Big(2^{\ft18} + \fft{280\, 2^{3/8}\, \epsilon}{3}\Big) \Big(
1 + (\fft1{2\, 2^{1/4}} - \fft{3080\epsilon}{3})\, t^2 + \cdots\Big)
\,,
\eea
and at large distances, they behaves as
\bea
a &=& 
\sqrt{\ft38}\, t\,\Big(1 + \ft43 (\ft23)^{2/3}\, t^{-\ft83} -
     \ft{80}{351}\, (\ft23)^{1/3}\, t^{-\ft{16}3} + 512\epsilon\, t^{-6}
+ \cdots\Big)\,,\nn\\
b&=&\sqrt{\ft38}\, t\,\Big(1 - \ft43 (\ft23)^{2/3}\, t^{-\ft83} -
     \ft{80}{351}\, (\ft23)^{1/3}\, t^{-\ft{16}3} + 512\epsilon\, t^{-6}
+ \cdots\Big)\,,\nn\\
c&=& \ft34 t\, \Big( 1 + \ft{320}{117} (\ft23)^{1/3}\, t^{-16/3} -
2048 \epsilon\, t^{-6} + \cdots\Big)\,.
\eea

\subsection{Corrections beyond ${\alpha'}^3$ order}

         The calculation for higher-order corrections up to order
$\alpha'^5$ is straightforward, but the results are rather complicated
to present in detail. We shall only list the large and small distance
behaviour in the comoving coordinate system.

\bigskip
\noindent\underline{$U(1)$ bundle over $S^2\times S^2\times S^2$}
\bigskip

        As in the previous case, we only consider the simplest case
with $a=b=c$.  We just give the large and small distance behaviour.  For
$r \rightarrow 0$, we have
\be
f=(\fft{1980\alpha'^3}{\ell^6} + \fft{10560\alpha'^4}{\ell^8} +
\fft{233520\alpha'^5}{\ell^{10}})\, \fft{\rho}{\ell^2} 
-(\fft{21018\alpha'^3}{\ell^6} + \fft{153312\alpha'^4}{\ell^8} +
\fft{4160920\alpha'^5}{\ell^{10}})\, \fft{\rho^2}{\ell^4} + \cdots
\,.
\ee
For $r\rightarrow \infty$, we have
\be
f=\fft{9\alpha'^3}{\rho^3} + \Big(\fft{69\alpha'^3}{\ell^6} +
\fft{3(1399+48\log(\rho/\ell^2))\alpha'^4}{16\ell^8} +
\fft{987520\alpha'^5}{273\ell^{10}}\Big)\, \fft{\ell^8}{\rho^4} +
\cdots\,.
\ee

\bigskip
\noindent\underline{$U(1)$ bundle over $S^2\times \CP^2$}
\bigskip

In this case, the higher-order correction to the function $s$ is given by
(assuming the simple case $a=c$)
\be
f=(\fft{1884\alpha'^3}{\ell^6} + \fft{95296\alpha'^4}{\ell^8} +
\fft{6298160\alpha'^5}{\ell^{10}})\, \fft{\rho}{\ell^2} 
-(\fft{20886\alpha'^3}{\ell^6} + \fft{4147168\alpha'^4}{\ell^8} +
\fft{37424680\alpha'^5}{\ell^{10}})\, \fft{\rho^2}{\ell^4} + \cdots
\,.
\ee
For $r\rightarrow \infty$, we have
\be
f=\fft{3\alpha'^3}{\rho^3} + \Big(\fft{87\alpha'^3}{\ell^6} +
\fft{38173+6568\log(\rho/\ell^2))\alpha'^4}{144\ell^8} +
\fft{8798080\alpha'^5}{2457\ell^{10}}\Big)\, \fft{\ell^8}{\rho^4} +
\cdots\,.
\ee

\bigskip
\noindent\underline{$U(1)$ bundle over $\CP^3$}
\bigskip

   In this case, the correction is easy to obtain, since we have
\be
S_3=\fft{495\alpha'^3\, X^3}{2a^6}\,,\qquad
S_4=\fft{4245\alpha'^4\, X^4}{4a^8}\,,\qquad
S_5=\fft{149145\alpha'^5\, X^5}{8a^{10}}\,,
\ee
where $X=(a^2-2g^2)/a^2$.  The perturbation function $f$ is given by
\bea
f&=& 45\alpha'^3\,\Big(\fft{2}{r^3} + \fft{2\ell^{8}}{r^{7}} +
\fft{2\ell^{16}}{r^{11}} + \fft{-11\ell^{24}}{2r^{15}} -
\fft{1}{\ell^4\,(r+\ell^2)} + 
\fft{45(r-\ell^2)}{\ell^4\,(r^2 +\ell^4)}\Big)\nn\\
&& \ft{4245}{16}\alpha'^4\, \Big(\fft{1}{r^4} + \fft{\ell^8}{r^8} +
\fft{\ell^{16}}{r^{12}} + \fft{\ell^{24}}{r^{16}} - 
\fft{4\ell^{32}}{r^{20}}\Big) + \ft{49715}{56}\alpha'^5\, \Big(
\fft{4}{\ell^8\, r} + \fft{4}{r^5}\nn\\
&&+\fft{4\ell^8}{r^9} + \fft{4\ell^{16}}{r^{13}} +
\fft{4\ell^{24}}{r^{17}} + \fft{4\ell^{32}}{r^{21}} -
\fft{21\ell^{40}}{r^{25}} -\fft{2}{\ell^8\, (r+\ell^2)} -
\fft{2(r+\ell^2)}{\ell^8\, (r^2 + \ell^4)}\Big)\,,\nn
\eea
where $r=\rho-\ell^2$.

\bigskip
\noindent\underline{Stenzel metric}
\bigskip

       The structure in this case is again rather complex.  We shall only
present $S_3$, $S_4$ and $S_5$, which are given by
\bea
S_3&=&-\fft{80e^{\ft{45}2 r}}{3^{1/4} (1 + e^{2r})^{15}\,
(1 + 4 e^{2r} + e^{4r})^{15/4}}\,\Big(51096822 + 82052992 \cosh{2 r}
\nn\\
&& +43709116 \cosh{4 r} +
16111758 \cosh{6 r} + 4256544 \cosh{8 r} + 823522 \cosh{10 r}\nn\\
&& +117252 \cosh{12 r} + 12021 \cosh{14 r} + 
810 \cosh{16 r} + 27 \cosh{18 r}
\Big)\nn\\
S_4&=&-\fft{1}{37748736 \cosh^{20}r\,(2 + \cosh{2 r})^5}\Big(
1912969150222 + 3256894409584 \cosh{2 r}\nn\\
&& + 2031561604552 \cosh{4 r} +
949210599696 \cosh{6 r} + 339746197983 \cosh{8 r}\nn\\
&&+94849519304 \cosh{10 r} + 20896666964 \cosh{12 r} +
3648186232 \cosh{14 r}\nn\\
&& + 501788290 \cosh{16 r} + 53212824 \cosh{18 r} +
4135140 \cosh{20 r} + 210600 \cosh{22 r}\nn\\
&& + 5265 \cosh{24 r}\Big)\nn\\
S_5 &=& -\fft{160e^{\ft{75}2r}}{9\, 3^{3/4}\, (1 + e^{2r})^{25}\,
(1 + 4e^{2r} + e^{4r})^{25/4}}\, \Big(
3192260095227860\nn\\
&& + 5621631779278675 \cosh{2 r} +
3857631050654430 \cosh{4 r}\nn\\
&& + 2088920390620110 \cosh{6 r} +
906177970153450 \cosh{8 r}\nn\\
&& + 319423127230328 \cosh{10 r} +
92531981651010 \cosh{12 r} + 22191184070115 \cosh{14 r}\nn\\
&& + 4418078655500 \cosh{16 r} + 728378998045 \cosh{18 r} +
98534694870 \cosh{20 r}\nn\\
&& + 10738276020 \cosh{22 r} + 912024630 \cosh{24 r} +
56882790 \cosh{26 r}\nn\\
&& + 2320650 \cosh{28 r} + 46413 \cosh{30 r}\Big)
\eea
 From these, it is straightforward to find the perturbation functions
$f$ and $g$, given in (\ref{fgsols}).

\section{Conclusion}\label{concl}

In this paper, we have shown how the preservation of supersymmetry on
BPS backgrounds such as non-compact Calabi-Yau spaces can be used to
obtain explicit expressions for the string-theory-derived $\alpha'$
corrections to these backgrounds. The corrected Killing spinor
conditions are the key to this. Even in the absence of full knowledge
of the supersymmetric structure of the $\alpha'$ corrections, these
Killing spinor conditions can be deduced from the requirement that the
corrected bosonic effective field equations appear as integrability
conditions for them. It is to be hoped that these corrected conditions
may illuminate the problem of supersymmetrising the string-theory
corrections, and in particular the important quartic curvature
corrections arising at order ${\alpha'}^3$, for which partial results
have been given in \cite{ggp,gsethi,pvhw}.

For K\"ahler manifolds, the scheme adopted in this paper has the
virtue of preserving the K\"ahler structure, although the
Ricci-flatness of the space is necessarily lost, since the deformed space
develops a new $U(1)$ factor in its holonomy. At the same time, the
dilaton $\phi$ acquires corrections as given in Eqn\
(\ref{dilcorr}). This latter point is of little significance, since it
can clearly be reset at order $\alp$ by defining a new dilaton $\td\phi$ 
that is related to $\phi$ by
\be
\td\phi = \phi + \ft12 \alp\, S_3\,.\label{phiredef}
\ee
A consequence of this redefinition is merely to change the specific
form of the $\alp$ corrections in the effective Lagrangian.  One of these
changes is a modification of the coefficient of $Y_2^\2$ in 
(\ref{ycomplete}).  Moreover, as noted previously, this coefficient 
can be adjusted by field redefinitions in a sigma-model calculation of
the effective action.  If one wants to avoid altering this coefficient,
one can achieve this by making a compensating transformation of the 
metric; for example by sending $g_{ab}\longrightarrow \td g_{ab}$ with
\be
\tilde g_{ab}= e^{-{1\over8}{\alpha'}^3\, S_3}\, g_{ab}
\ee
at the same time as $\phi\longrightarrow \td \phi$
The change in the Ricci tensor under a Weyl transformation
$g_{ab}\rightarrow \tilde g_{ab}=e^{2\sigma}g_{ab}$ in a space of
dimension $D$ is
\be
\tilde R_{ab}=R_{ab} -
(D-2)\nabla_a\nabla_b\sigma-(D-1)(D-2)
\nabla_a\sigma\nabla_b\sigma - \square\sigma
g_{ab}\ ,
\ee
so, setting $D=10$ and keeping terms only to order ${\alpha'}^3$,
the corrected Einstein equation (\ref{ricmod}) becomes
\be
\wtd R_{ij}={\alpha'}^3(\wtd\nabla_{\hat i}\wtd\nabla_{\hat j}\wtd S_3-
\ft18\wtd\square \wtd S_3\,\td g_{ij})\label{modboseq}
\ee
in the conformally-related metric.  The price to be paid for doing this
is that the metric $\td g_{ij}$ is no longer K\"ahler.

   A similar type of field redefinition, but expressible in terms of a
purely six-dimensional Weyl scaling, takes one from the
K\"ahler-preserving scheme employed in this paper to the scheme used
in \cite{tseytlin}.  From a geometrical point of view a scheme that
preserves the K\"ahler structure of the metric is appealing.  Schemes
that do not preserve the K\"ahler structure would appear to have a
more {\it ad hoc} character.

    The technique for obtaining explicit expressions for $\alpha'$
corrections to internal manifolds employed in this paper extends
naturally to $D=7$ manifolds with $G_2$ holonomy. This is discussed
separately in Ref.\ \cite{g2paper}.

\section*{Acknowledgments}

We would like to thank Dominic Joyce, Kasper Peeters, Paul Townsend, 
Arkady Tseytlin and  Pierre Van Hove for helpful discussions and 
correspondence. For
hospitality at various stages during the course of the work,
K.S.S. would like to thank Texas A\&M University and C.N.P. and
K.S.S. would like to thank the Benasque Center for Science and the
Universities of Barcelona, Cambridge and Groningen.

\newpage

\centerline{\Large{\bf APPENDICES}}
\appendix

\section{Topological Invariants and the {\it Curvatura Integra}}
\label{curvaturasec}

   The Euler number of a compact manifold $M$ of (even) dimension 
$n=2p$ is given by integrating the $n$ form 
\be
\Psi \equiv \fft{1}{p!\, (4\pi)^p}\, \ep^{a_1b_1\cdots a_p b_b}\, 
   \Theta_{a_1 b_1} \wedge \cdots \wedge \Theta_{a_p b_p}\,,
\ee
where $\Theta_{ab}= d\omega_{ab} + \omega_a{}^c\wedge \omega_{cb}$ is 
the curvature 2-form; $\chi = \int_M \Psi$.   
The $n$-form $\Psi$ can be rewritten as $\Psi= E_n\, \sqrt{g}\, d^nx$, 
where the ``Euler integrand'' $E_n$ is given by
\be
E_n = \fft{(2p-1)!!}{(4\pi)^p}\,  R_{a_1 a_2}{}^{[a_1 a_2}\, 
     R_{a_3 a_4}{}^{a_3 a_4}\cdots   R_{a_{n-1} a_n}{}^{a_{n-1} a_n]}\,.
\ee

   In a non-compact manifold, the Euler number is not given just by the
volume integral of the Euler integrand; there is also a boundary term
that must be included \cite{chern}:
\be
 \chi= \int_M \Psi + \int_{\del M} \Phi\,,\label{nceuler}
\ee
where in $n=2p$ dimensions the {\it Curvatura Integra} $\Phi$ is an
$(n-1)$-form constructed from the Riemann curvature and the second
fundamental form of the boundary.  It is shown in \cite{chern} that if
$u^a$ denotes the unit outward-pointing vector normal to the boundary,
then $\Phi$ is given by
\be
\Phi= \fft1{(2\pi)^p}\, \sum_{m=0}^{p-1} \fft{2^{-m}}{m!\, (2p-2m-1)!!}\, 
        \Phi_{(m)}\,,\label{boundaryphi}
\ee
where 
\be
\Phi_{(m)} = \ep^{a b_1\cdots b_{n-2m-1} c_1 d_1 \cdots c_m d_m}\, 
    u_a \, \theta_{b_1}\wedge \cdots \wedge \theta_{b_{n-2m-1}}\wedge
\Theta_{c_1 d_1}\wedge \cdots \wedge \Theta_{c_m d_m}\,.\label{phim}
\ee
The second fundamental form is defined by
\be
\theta_a = Du_a \equiv du_a + \omega_{ab}\, u^b\,.
\ee

   In the case of metrics $ds^2 = dt^2 + d\bar s^2(t)$, which includes
all our examples in sections \ref{cy6sec} and \ref{cy8sec}, the unit
vector normal to the boundary at $t=t_0$ is just given by $u=\del/\del
t$, and so we shall have $u_0=1$, $u_i=0$ for $i\ge 1$.  Thus we have
\be
\theta_0=0\,,\qquad \theta_i= -\omega_{0i}\,,\quad i\ge 1\,.
\ee

    In six dimensions, equation (\ref{boundaryphi}) gives
\be
\Phi= \fft1{8\pi^3}\, \Big[ \ft1{15}\, \Phi_{(0)} + \ft16 \Phi_{(1)} + 
\ft18 \Phi_{(2)}\Big]\,,
\ee
and (\ref{phim}) gives
\bea
\Phi_{(0)} &=& \ep^{ijk\ell m}\, \theta_i\wedge \theta_j\wedge \theta_k
\wedge \theta_\ell \wedge \theta_{m}\,,
\nn\\
\Phi_{(1)} &=& \ep^{ijk\ell m}\, \theta_i\wedge \theta_j\wedge\theta_k
\wedge \Theta_{\ell m}\,,
\nn\\
\Phi_{(2)} &=& \ep^{ijk\ell m}\, \theta_i\wedge \Theta_{jk}
\wedge \Theta_{\ell m}\,.
\eea

   In eight dimensions, the corresponding expressions are given by
\be
\Phi= \fft1{16\pi^4}\, \Big[ \ft1{105}\, \Phi_{(0)} + \ft1{30} \Phi_{(1)} + 
\ft1{24} \Phi_{(2)} + \ft1{48} \Phi_{(3)}\Big]\,,
\ee
with
\bea
\Phi_{(0)} &=& \ep^{ijk\ell mpq}\, \theta_i\wedge \theta_j\wedge \theta_k
\wedge \theta_\ell \wedge \theta_{m}\wedge \theta_p\wedge \theta_q\,,
\nn\\
\Phi_{(1)} &=& \ep^{ijk\ell mpq}\, \theta_i\wedge \theta_j\wedge\theta_k
\wedge \theta_\ell\wedge\theta_m\wedge \Theta_{pq}\,,
\nn\\
\Phi_{(2)} &=& \ep^{ijk\ell mpq}\, \theta_i\wedge \theta_j\wedge\theta_k
\wedge \Theta_{\ell m}
\wedge \Theta_{pq}\,,\nn\\
\Phi_{(3)} &=& \ep^{ijk\ell mpq}\, \theta_i\wedge \Theta_{jk}
\wedge \Theta_{\ell m}
\wedge \Theta_{pq}\,.
\eea

     It is interesting to note that if we vary the metric $g_{ab}$ in
$E_8$ then those terms linear in $R_{ab}$ are given by
\be
\delta E_8  
= \fft{1}{4\pi}\, E_6\, R_{ab}\, \delta g^{ab} 
           \,,\label{ricciterms}
\ee
where $E_6$ is precisely the Euler integrand of six dimensions
(including all Ricci-tensor terms),
\be
E_6 = \fft{15}{(4\pi)^3}\, R_{a_1 a_2}{}^{[a_1 a_2}\, 
              R_{a_3 a_4}{}^{a_3 a_4}\, R_{a_5 a_6}{}^{a_5 a_6]}\,.
\ee

   The cubic curvature invariant
$S_3$ given in (\ref{s3exp}) is proportional, modulo terms involving the
Ricci tensor, to the Euler integrand $E_6$ in six dimensions.  The 
exact expression for $E_6$, including all Ricci terms, is
\bea
E_6 &\equiv&  \fft{15}{64\pi^3}\, R_{a_1 a_2}{}^{[a_1 a_2}\, 
     R_{a_3 a_4}{}^{a_3 a_4}\,  R_{a_5 a_6}{}^{a_5 a_6]}\,,\nn\\
&=& 
\fft{S_3}{96 \pi^3}+\fft1{384\pi^3}\, (-24 R_{abcd}\, R^{abc}{}_e\, R^{de} 
    + 3R\, R_{abcd}\, R^{abcd} + 24 R^{abcd}\, R_{ac}\, R_{bd}\nn\\
&&\qquad\qquad\qquad\quad
 + 16 R_a{}^b\, R_b{}^c\, R_c{}^a - 12 R\, R_{ab}\, R^{ab} + R^3)\,.
\eea
Thus we see that when evaluated in the Ricci-flat unperturbed Calabi-Yau
metric, we shall have
\be
\fft1{96\pi^3}\, \int_M S_3\, \sqrt{g}\, d^6x \equiv 
\wtd\chi= (\chi - \Xi)\,,\qquad
\hbox{where}\quad \Xi\equiv \int_{\del M} \Phi\,.
\ee
Here $\Xi$ is the contribution to the Euler number from the surface
term in (\ref{nceuler}).  Thus the quantity 
$\wtd\chi$ that results from integrating
$S_3$ over a non-compact Calabi-Yau manifold is neither the Euler number
nor is it a topological invariant.

\section{Euler Numbers from Curvature Integrals}

   In Appendix A, we review some standard material on the calculation
of the Euler number in terms of integrals over quantities formed from
the curvature of the metrics.  Because the manifolds $M$ that we are
studying here are non-compact, it is necessary to include the
contributions not only of the usual volume term in the Euler
integrand, but also a contribution coming from the boundary $\del M$
that one can introduce in order to compactify the manifold. 
The Euler number is then given by
\be
\chi = \int_{M_0} \Psi   + \int_{\del M_0} \Phi\,, \label{oceuler}
\ee
where $M_0$ denotes the compact manifold introduced by cutting off the
$n$-dimensional non-compact manifold $M$ with a boundary $\del M_0$.
The answer is, of course, independent of any smooth deformation of
$\del M_0$.  It is useful to introduce the notation $\del M$ to denote
the limiting case where the boundary is pushed out all the way to 
infinity.  The $n$-form $\Psi$ is the usual Euler form, and the 
$(n-1)$-form $\Phi$ is the {\it curvatura integra} that is constructed
in \cite{chern}, which supplies the boundary term.

   It is now a mechanical exercise to calculate the contributions given in
(\ref{oceuler}) to the Euler number for each of the manifolds we have
considered here.  Considering first the six-dimensional cases in section 
\ref{cy6sec}, we find
\bea
\hbox{\bf Resolved Conifold}:&& \int_M \Psi = \ft{14}{27}\,,\qquad
                      \int_{\del M}  \Phi= \ft{40}{27}\,,\qquad 
                 \chi=\ft{14}{27} + \ft{40}{27} = 2\,,\nn\\
\hbox{\bf Deformed Conifold}:&& \int_M \Psi = -\ft{40}{27}\,,\qquad
                      \int_{\del M}  \Phi= \ft{40}{27}\,,\qquad 
                 \chi=-\ft{40}{27} + \ft{40}{27} = 0\,,\nn\\
\hbox{\bf $\R^2$ bundle over $S^2\times S^2$}:&& 
            \int_M \Psi = \ft{88}{27}\,,\qquad
                      \int_{\del M}  \Phi= \ft{20}{27}\,,\qquad 
                 \chi=\ft{88}{27} + \ft{20}{27} = 4\,,\nn\\
\hbox{\bf $\R^2$ bundle over $\CP^2$}:&& 
            \int_M \Psi = \ft{8}{3}\,,\qquad
                      \int_{\del M}  \Phi= \ft{1}{3}\,,\qquad 
                 \chi=\ft{8}{3} + \ft{1}{3} = 3\,,
\eea
where the boundary is taken to be at $t=t_0$, in the limit where 
$t_0\longrightarrow\infty$. 
These results are all consistent with expectation.  The resolved conifold
is an $\R^4$ bundle over $S^2$, whose Euler number is the same as
that for a direct product $\R^4\times S^2$, giving $\chi= 1\times 2=2$.
The deformed conifold is an $\R^3$ bundle over $S^3$, giving $\chi=
1\times 0=0$.  The $\R^2$ bundles over $S^2\times S^2$ and $\CP^2$
give $\chi= 1\times 2\times 2=4$ and $\chi=1\times 3=3$ respectively.

   It should be noted that even in a case such as the deformed conifold,
which has zero Euler number, the volume integral of the Euler integrand
$E_6$ is non-zero. 

    We now turn to the eight-dimensional metrics that we considered
in section \ref{cy8sec}.  For these, we find
\bea
\hbox{\bf $\R^2$ bundle over $S^2\times S^2\times S^2$}:
&& \int_M \Psi = \ft{111}{16}\,,\qquad
                      \int_{\del M}  \Phi= \ft{17}{16}\,,\qquad 
                 \chi= 8\,,\nn\\
\hbox{\bf $\R^4$ bundle over $S^2\times S^2$}:
&& \int_M \Psi = \ft{15}{8}\,,\qquad
                      \int_{\del M}  \Phi= \ft{17}{8}\,,\qquad 
                 \chi= 4\,,\nn\\
\hbox{\bf $\R^2$ bundle over $S^2\times \CP^2$}:
&&            \int_M \Psi = \ft{687}{128}\,,\qquad
                      \int_{\del M}  \Phi= \ft{81}{128}\,,\qquad 
                 \chi= 6\,,\nn\\
\hbox{\bf $\R^4$ bundle over $\CP^2$}:&& 
            \int_M \Psi = \ft{111}{64}\,,\qquad
                      \int_{\del M}  \Phi= \ft{81}{64}\,,\qquad 
                 \chi= 3\,,\nn\\
\hbox{\bf $\R^2$ bundle over $\CP^3$}:&& 
            \int_M \Psi = \ft{15}{4}\,,\qquad
                      \int_{\del M}  \Phi= \ft{1}{4}\,,\qquad 
                 \chi= 4\,,
\eea
Again, these Euler numbers accord with one's expectations, since the Euler
number for a fibre $\R^m$ over a base $B$ is just given by the Euler
number of $B$, and we know that $\chi(S^2)=2$, $\chi(\CP^2)=3$ and
$\chi(\CP^3)=4$.   

    Note that although one customarily tends to
evaluate the volume and boundary contributions to the Euler number
by choosing a boundary surface that is pushed out to infinity, as in 
our results presented above, the boundary can equally well be chosen to
be at any radius.  We have explicitly verified for all the six-dimensional
and eight-dimensional examples listed above that one indeed gets the 
identical results for $\int_M \Psi + \int_{\del M} \Phi$ when the
bounding surface is taken to be at any radius $r_0$.  This provides a
useful check that the computations of $\Psi$ and $\Phi$, which are 
quite involved, are indeed correct. 

   An interesting limiting choice for the radius of the bounding surface
is to take it to lie at $r_0=0$; \ie at the origin, on the base $B$ of the
$\R^n$ fibre bundle over $B$.  In this case, there is no contribution at
all from the volume integral $\int_M \Psi$, and the entire contribution
to the Euler number comes from the boundary term $\int_{\del M} \Phi$, 
with $\Phi$ evaluated at $r=0$.

\end{document}